# THE THEORY OF SURFACE ENHANCED HYPER RAMAN SCATTERING

[1]V.P. Chelibanov [2]A.M. Polubotko

[1]State University of Information Technologies, Mechanics and Optics, Kronverkskii 49, 197101 Saint Petersburg, RUSSIA  E-mail: Chelibanov@gmail.com

[2]A.F. Ioffe Physico-Technical Institute, Politechnicheskaya 26, 194021 Saint Petersburg Russia

Tel: (812) 274-77-29, Fax: (812) 297-10-17, E-mail: alex.marina@mail.ioffe.ru

## ABSTRACT

The dipole-quadrupole theory of Surface Enhanced Hyper Raman scattering (SEHRS), created by the authors is expounded in detail. Peculiarities of the behavior of electromagnetic field on rough metal surfaces are considered. It is demonstrated that there is an enhancement of the dipole and quadrupole light-molecule interaction near the places with a large curvature in surface fields, strongly varying in space, which exist in the system. The quantum mechanical description of molecule wavefunctions and the cross-section of the enhanced Hyper Raman scattering exposed in detail. It is demonstrated that the cross-section consists from several contributions, which describe the scattering via various dipole and quadrupole moments. Then selection rules for the scattering contributions are obtained and a qualitative classification of the contributions after the enhancement degree is performed. Analysis of experimental spectra of pyrazine and phenazine, and also some another molecules is performed too. It is demonstrated full coincidence of experimental regularities in these spectra with the theory suggested. In particular it concerns the effect of appearance of strong forbidden lines, caused by totally symmetric vibrations with the unit irreducible representation. In addition it is demonstrated that the strong quadrupole light-molecule interaction is forbidden in molecules with cubic symmetry groups, such as methane, due to so-called Electrodynamical forbiddance, which is a consequence both of the electrodynamical laws and the cubic symmetry of the molecules. Therefore the main peculiarity of the SEHR spectra is that the most enhanced lines in such molecules are caused by another irreducible representations then the unit one. In addition the nature of the first layer effect in SEHRS is considered. It is demonstrated that this effect is associated with the strong change of the electric field and its derivatives in the first and the second layer, but not with "the chemical enhancement". Peculiarities of the SEHRS frequency dependence are considered too.

## 1. INTRODUCTION

The enhanced optical processes on molecules, adsorbed on rough metal surfaces were discovered in 1974 [1] with the discovery of SERS. The essence of this phenomenon is that the molecules, adsorbed on such surfaces enhance the intensity of the Raman scattering lines, and the enhancement coefficient is ~ $10^6$. Further another surface enhanced optical processes, such as an enhancement of infrared absorption [2], the enhancement of second harmonic generation [3], and also the enhancement of Hyper Raman scattering [4-6] were discovered too. This work is devoted for creation of the theory just of the last phenomenon. There are few works, which are devoted for creation of the theory of SEHRS. They are the papers [7-9] and also our papers [10-14]. In [7-9] and in majority of experimental works, the ideas about surface plasmons and also about "chemical mechanism", which is associated with the direct touch of the molecule with the surface are used for interpretation of experimental results. From our point of view, both these mechanisms are mistakable. We consider that the reason of SEHRS and of the enhancement of other optical processes is so-called rod effect or the enhancement of the electric field near the spike, or tip. In addition one must take into account both the dipole and quadrupole light-molecule interaction. Appearance of strong forbidden lines in the spectra of such symmetrical molecules as pyrazine and phenazine [15-17] supports our point of view. In case, when only the dipole interaction is taken into account, the appearance of these lines can not be explained in principle. However, when the quadrupole interaction is taken into account, this fact expands the number of moments, which form the spectrum and the appearance of the forbidden lines can be explained naturally. Further we shall expound the dipole-quadrupole theory of SEHRS and then we shall comment the most essential literature on some experimental works on SEHRS.

Our approach to explanation of SEHRS essentially differs from another approaches since the quadrupole interaction in surface fields strongly varying in space is taken into account.

The plasmon hypothesis is based on the fact that there are some solutions of Maxwell equations on flat and sphere surfaces of a plasmon type. This type of solutions have some peculiarities of a resonance type when $\varepsilon' = -1$ and $\varepsilon' = -2$ respectively. Here $\varepsilon'$ is a real part of the dielectric constant of the metal. One should note that even in these both cases the position of the plasmon resonances depends on the geometry. Therefore in order to speak about plasmons on random rough surface, where SEHRS is observed, it is necessary to solve a diffraction problem on such a surface. However it is impossible, since the problems of mathematical physics can be analytically solved only for several coordinate systems. Therefore the idea of plasmons on random rough surface is not defined. Further discussion about plasmons is uselessly. Moreover there is no such theories in literature on the enhanced optical processes.

As it was pointed out, the second mechanism, which is often cited in literature is the "chemical mechanism", which associates the enhancement with the touch of the molecule with the substrate and therefore manifests only in the first layer of adsorbed molecules. From our point of view this idea is mistakable also.

Below we shall demonstrate that strong difference of the enhancement in the first and the second layers of adsorbed molecules is associated with a very large difference in the values of the electric field (and its derivatives) that results in the strong and primary enhancement of the processes in the first layer [18,19].

From our point of view the reason of SEHRS is so called rod effect or the enhancement of the electric field near the spike or tip. Moreover one must take into account not only the dipole, but the quadrupole light-molecule interaction. Here we shall state the dipole-quadrupole theory of SEHRS, published in [10-14] and further we shall comment some literature on experimental works in this area, which is of interest for corroboration of the theory.

## 2. MAIN PROPERTIES OF THE ELECTROMAGNETIC FIELD NEAR ROUGH METAL SURFACE

The main property of electromagnetic field near rough metal surface is its strong spatial heterogeneity. As an example we can consider a model of a rough surface - a strongly jagged metal lattice with a regular triangular profile of the eshelett type (Figure 1). The electromagnetic field above this lattice can be represented in the form

$$\overline{E} = \overline{E}_{inc} + \overline{E}_{surf..sc} \tag{1}$$

where,

$$\overline{E}_{inc} = \overline{E}_{0,inc} e^{-ik_0 \cos\theta_0 z + ik_0 \sin\theta_0 y}$$
$$\left|\overline{E}_{0,inc}\right| = 1 \tag{2}$$

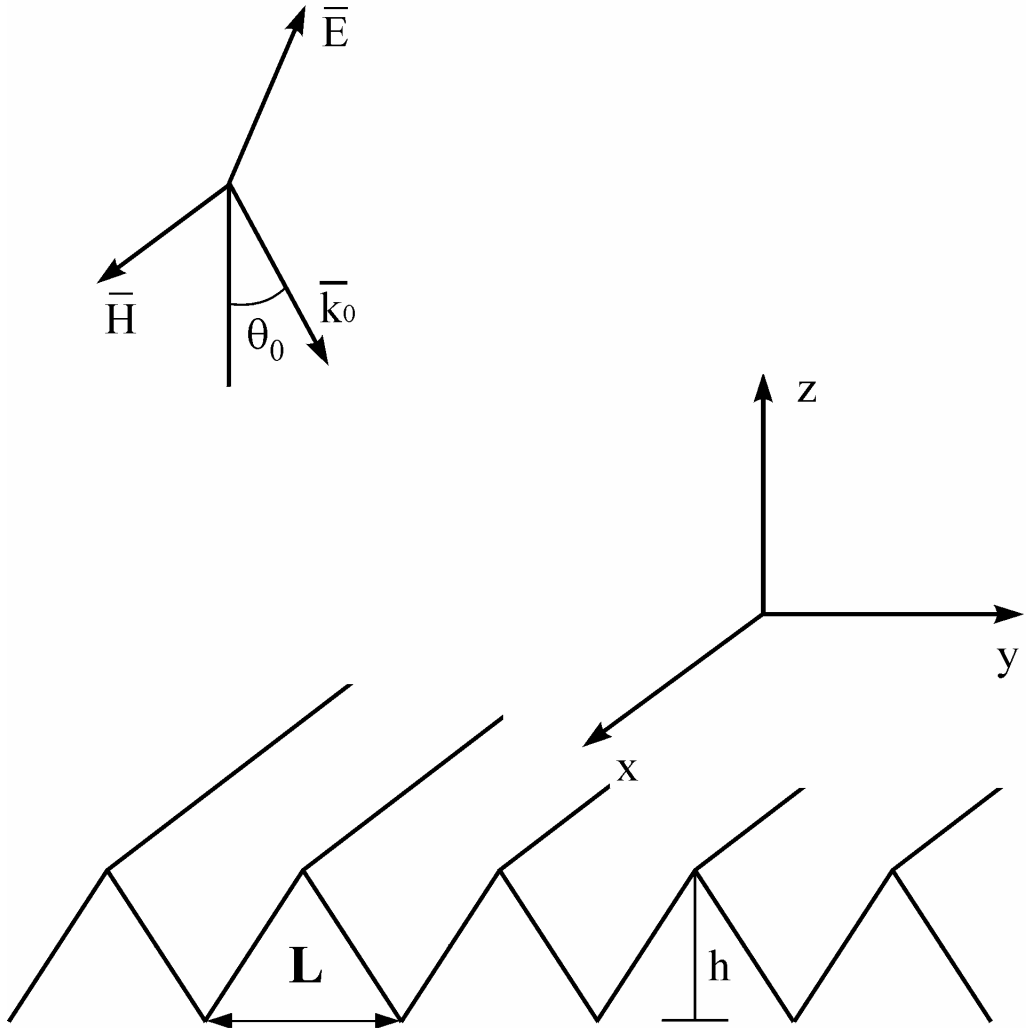

Figure 1. Diffraction of the plane $H$ polarized wave on the lattice of the eshelett type. Here $L \ll \lambda$, $L$ is a lattice period, $h$ is a height of the lattice.

$\theta_0$ - is the angle of incidence, $\bar{k}_0$ is the wave vector of the incident field in a free space,

$$\bar{E}_{surf.sc.} = \sum_{n=-\infty}^{+\infty} \bar{g}_n e^{i\alpha_n y + i\gamma_n z} \qquad (3)$$

$$\alpha_n = \frac{2\pi n + k_0 \sin\theta_0 L}{L} \qquad (4)$$

$$\gamma_n = \sqrt{k_0^2 - \alpha_n^2} \qquad (5)$$

Here $\bar{g}_n$ are the amplitudes, $n$ is the number of a spatial harmonic. For the period of the lattice $L \ll \lambda$ the spatial harmonic with $n = 0$ is a direct reflected wave, while all others are heterogeneous plane waves strongly localized near the surface. The maximum localization size has the harmonic with $n = 1$. All others are localized considerably stronger. The exact solution of the diffraction problem on the lattice reduces to determination of the coefficients $\bar{g}_n$. The main specific feature of the surface field is a steep or singular increase of the electric field near the wedges of the lattice or so-called rod effect. This type of behavior is independent on a particular surface profile. It is determined only by existence of sharp wedges. Besides it is independent on the dielectric properties of the lattice and exists in lattices with any dielectric constants that differ from the dielectric constant of vacuum.

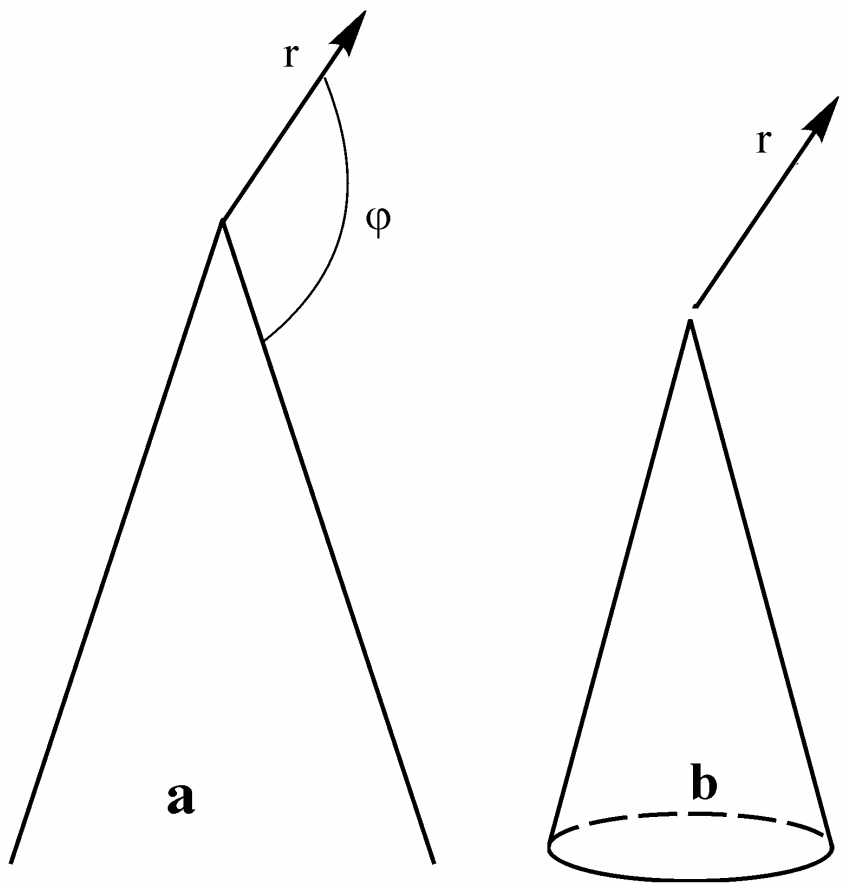

Figure 2. **a**-infinite wedge, **b**-the roughness of the cone type.

In the vicinity of the wedge (Figure 2a) the electric field can be estimated as

$$E_r = -g_{0,inc} C_0 \left(\frac{l_1}{r}\right)^\beta \sin(\lambda_1 \varphi)$$
$$E_\varphi = g_{0,inc} C_0 \left(\frac{l_1}{r}\right)^\beta \cos(\lambda_1 \varphi)$$
(6)

where $C_0$ is some numerical coefficient, ($l_1 = L$ or $h$) is a characteristic size of the lattice

$$\lambda_1 = \pi/(2\pi - \alpha) \qquad (7)$$

$\alpha$ is the wedge angle. For an ideally conductive wedge.

$$\beta = 1 - \lambda_1 = \frac{\pi - \alpha}{2\pi - \alpha} \qquad (8)$$

The specific feature of the field behavior (6) is appearance of the singularity $(l_1/r)^\beta$, which describes geometrical nature of the field enhancement. It determines the following behavior of the coefficients $\overline{g}_n$ in expression (3)

$$g_n \sim |n|^{\beta-1}. \qquad (9)$$

Indeed, substitution of (9) into (3) gives

$$\sum_{\substack{n=-\infty \\ n \neq 0}}^{+\infty} |n|^{\beta-1} e^{2\pi|n|z/L} \sim 2\int_0^\infty t^{\beta-1} e^{-2\pi z t/L} dt \sim 2\left(\frac{L}{2\pi z}\right)^\beta \qquad (10)$$

For the wedge angles changing in the interval $0 < \alpha < \pi$ the $\beta$ value varies within the interval $0 < \beta < 1/2$ and the coefficients $g_n$ slowly decrease as $n$ increases. Thus the singular behavior of the field arises because of specific summation of the surface waves at the top of the wedge. In the region of a three-dimensional roughness of the cone type (Figure 2b) the formula for estimation of the field has an approximate form

$$E_r \sim g_{0,inc} C_0 \left(\frac{l_1}{r}\right)^\beta, \qquad (11)$$

here $\beta$ depends on the cone angle and varies within the interval $0 < \beta < 1$. Using formulae (6)

and (11) one can note a very important property: a strong spatial variation of the field. For example

$$\frac{1}{E_r}\frac{\partial E_r}{\partial r} \sim \left(\frac{\beta}{r}\right) \qquad (12)$$

can be significantly larger than the value $2\pi/\lambda$, which characterizes variation of the electric field in a free space. If one considers more realistic models of the rough surface than the regular metal lattice, it is obvious, that there is a strong enhancement of the perpendicular component of the electric field $\overline{E}_n$ at places with a large curvature (Figure 3), while the tangential components $\overline{E}_\tau$ are comparable with the amplitude of the incident field.

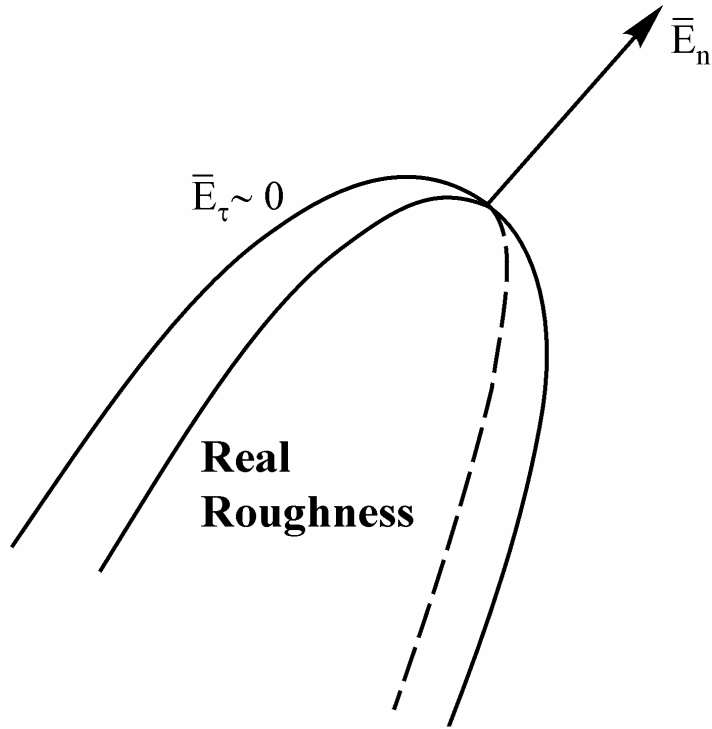

Figure 3. More realistic model of the roughness of the cone type. The normal component of the electric field $\overline{E}_n$ and the derivatives $\partial E_\alpha/\partial x_\alpha, \alpha = (x, y, z)$ are enhanced near the top of the roughness.

Besides, the electromagnetic field strongly varies in space with a characteristic length $l_E$ equal to characteristic roughness size. This type of behavior is not an exclusive property of the ideally conductive lattice and preserves near surfaces with a finite dielectric constant.

## 3. INTERACTION OF LIGHT WITH MOLECULES NEAR ROUGH METAL SURFACE

In accordance with principles of theoretical physics the optical properties of molecules are determined by the light-molecule interaction Hamiltonian, which has the form

$$\widehat{H}_{e-r} = -\sum_i \frac{ie\hbar}{mc} \overline{A}_i \overline{\nabla}_i \quad . \tag{13}$$

Here the sign of $e$ is positive. $\overline{A}_i$ is a vector potential of the electromagnetic field at the place of the $i$ electron. Other designations are conventional. For small objects, like molecules, the vector potential can be expanded in a Taylor series and the final expressions for the light-molecule interaction Hamiltonians for the incident and scattered fields can be obtained in the form

$$\widehat{H}^{inc}_{e-r} = \left|\overline{E}_{inc}\right| \frac{(\overline{e}^* \overline{f}_e^*)_{inc} e^{i\omega_{inc} t} + (\overline{e}\,\overline{f}_e)_{inc} e^{-i\omega_{inc} t}}{2} \tag{14}$$

$$\widehat{H}^{scat}_{e-r} = \left|\overline{E}_{scat}\right| \frac{(\overline{e}^* \overline{f}_e^*)_{scat} e^{i\omega_{scat} t} + (\overline{e}\,\overline{f}_e)_{scat} e^{-i\omega_{scat} t}}{2} \tag{15}$$

where $\overline{E}_{inc}$ and $\overline{E}_{scat}$ are vectors of the incident and scattered electric fields, $\omega_{inc}$ and $\omega_{scat}$ are corresponding frequencies, $\overline{e}$ - are polarization vectors of the corresponding fields,

$$f_{e\alpha} = d_{e\alpha} + \frac{1}{2E_\alpha} \sum_\beta \frac{\partial E_\alpha}{\partial x_\beta} Q_{e\alpha\beta} \tag{16}$$

is an $\alpha$ component of the generalized vector of interaction of light with molecule,

$$d_{e\alpha} = \sum_i e x_{i\alpha} \quad Q_{e\alpha\beta} = \sum_i e x_{i\alpha} x_{i\beta} \tag{17}$$

are the $\alpha$ component of the dipole moment vector and the $\alpha\beta$ component of the quadrupole moments tensor of electrons of the molecule. Here under $x_{i\alpha}$ and $x_{i\beta}$ we mean coordinates $x, y, z$ of the $i$ electron. Usually the relative influence of the quadrupole and dipole light-molecule interactions is determined as the relation of the second and the first terms in the right side of (16). However this relation is the relation of quantum-mechanical operators, while all physical values are expressed via the matrix elements of these operators. Therefore one must consider, that the relative influence is determined by the relations

$$\frac{\overline{\langle m|Q_{e\alpha\beta}|n\rangle}}{\overline{\langle m|d_{e\alpha}|n\rangle}} \frac{1}{2E_\alpha} \frac{\partial E_\alpha}{\partial x_\beta} = B_{\alpha\beta} a \frac{1}{2E_\alpha} \frac{\partial E_\alpha}{\partial x_\beta} \qquad (18)$$

where $a$ is a molecule size, $B_{\alpha\beta}$ are some numerical coefficients. The first factor in the left side of (18) is the relation of some mean matrix elements of the quadrupole and dipole transitions. It is necessary to point out that the $B_{\alpha\beta}$ values essentially differ for $\alpha \neq \beta$ and for $\alpha = \beta$. This fact results from the fact that $d_{e\alpha}$ and $Q_{e\alpha\beta}$ ($\alpha \neq \beta$) are the values with a changeable sign, while $Q_{e\alpha\alpha}$ are the values with a constant sign, that strongly increases the $B_{\alpha\alpha}$ values. The difficulty for this estimation is the fact, that there is no information about quadrupole transitions in molecules. Therefore because the inner shell configuration in molecules remains almost intact we usually take the value $\overline{\langle n|Q_{e\alpha\alpha}|n\rangle}$ instead of $\overline{\langle m|Q_{e\alpha\alpha}|n\rangle}$ and the value $\sqrt{e^2\hbar/2m\omega_{mn}} \times \sqrt{\overline{f}_{mn}}$ for $\overline{\langle m|d_\alpha|n\rangle}$, which is expressed in terms of some mean value of the oscillator strength $\overline{f}_{mn} = 0.1$, while $\omega_{mn}$ corresponds to the edge of absorption ($\lambda \approx 2500 A$). Since configuration of the electron shell is close to configuration of nuclei the value $\overline{\langle n|Q_{\alpha\alpha}|n\rangle}$ was calculated as a $Q_{n\alpha\alpha}$ component of the quadrupole moments of nuclei. Estimation for the pyridine, benzene or pyrazine molecules gives the value $B_{\alpha\alpha} \sim 2\times 10^2$ that strongly differs from the value $B_{\alpha\alpha} \sim 1$ ($B_{\alpha\alpha} a \sim a$). The last one is usually used in literature as the relation of the quadrupole and dipole operators. The full estimation of the relation (18) near rough metal surface is made for the models of roughness of the wedge or the cone or spike types, which well approximate prominent features of the surface with a large curvature. Then using (12) one can obtain the following expression for

$$\frac{\overline{\langle m|Q_{e\alpha\alpha}|n\rangle}}{\langle m|d_{e\alpha}|n\rangle} \frac{1}{2E_\alpha} \frac{\partial E_\alpha}{\partial x_\alpha} = B_{\alpha\alpha} a \frac{\beta}{2r} \qquad (19)$$

It can be seen, that for $r < \frac{1}{2} B_{\alpha\alpha} a \beta$ the quadrupole interaction can be more than the dipole one. The enhancement of the electric field and the dipole interaction can be estimated as

$$G_{H_d} \sim C_0 \left(\frac{l_1}{r}\right)^\beta \qquad (20)$$

while the enhancement of the quadrupole interaction compared to the dipole interaction in a free space can be estimated as

$$G_{H_Q} \sim C_0 \beta \left(\frac{B_{\alpha\alpha}}{2}\right) \left(\frac{l_1}{r}\right)^\beta \left(\frac{a}{r}\right) . \qquad (21)$$

It can be seen, that for the reasonable values $C_0 \sim 1$, $l_1 \sim 10$ nm, $r \sim 1$ nm, $\beta \sim 1$ and the molecules like pyridine, benzene or pyrazine with $B_{\alpha\alpha} \sim 2\times 10^2$ the enhancement of the dipole interaction is $\sim 10$, while the enhancement of the quadrupole interaction is $\sim 10^2$.

## 4. THE ENHANCEMENT IN SEHRS

The enhancement in SEHRS such as in SERS can be caused both by the enhancement of the $E_z$ component of the electric field, which is perpendicular to the surface and by the enhancement of the field derivatives and the quadrupole interaction. Since SEHRS is the process of the third order the enhancement due to the dipole interaction is

$$G_d \sim C_0^6 \left(\frac{l_1}{r}\right)^{6\beta} \qquad (22)$$

and

$$G_Q \sim C_0^6 \beta^6 \left(\frac{B_{\alpha\alpha}}{2}\right)^6 \left(\frac{l_1}{r}\right)^{6\beta} \left(\frac{a}{r}\right)^6 \qquad (23)$$

due to the quadrupole interaction. For the values of the parameters pointed out above, the enhancement due to the purely dipole interaction is of the order of $10^6$ while the enhancement due to the quadrupole interaction $\sim 10^{12}$. One should note that the enhancement of the dipole and quadrupole interactions $G_{H_d}$ and $G_{H_Q}$ and especially the enhancement of SEHRS $G_Q$ can be very large. For example for some limited situations with the values of the parameters $C_0 \sim 1$, $B_{\alpha\alpha} \sim 2\times 10^2$, $r \sim 0.1 nm$, $\beta \sim 1$, $l_1 \sim 100 nm$ corresponding to the placement of the molecule on the top of the cone (tip or spike) the enhancement $G_Q$ in SEHRS can achieve $10^{30}$. As it was mentioned above the real situation is that most enhancement arises in the vicinity of some points associated with prominent places with a very large curvature. The mean enhancement is formed from the whole layer of adsorbed molecules and is significantly smaller than the maximum enhancement near these places due to averaging.

## 5. MAIN AND MINOR MOMENTS

In accordance with the previous consideration the most enhancement arises from the quadrupole interaction with the moments $Q_{xx}, Q_{yy}, Q_{zz}$ and the dipole interaction caused by the enhancement of the electric field, which is perpendicular to the surface. Depending on orientation of the molecule the dipole light molecule interaction with various components of the $d$ moments can be important for the enhancement. For example for the pyrazine molecule lying flatly on the surface (with the $d_y$ moment perpendicular to the plane of the molecule) only the $d_y$ moment is essential. For the molecule bounded by the nitrogen atom with the surface the essential $d$ moment is $d_z$, while for a large coverage with arbitrary orientation of the molecules with respect to the $E_z$ component of the electric field, all the $d$ moments can contribute to the scattering. (This situation arises due to possible physisorption of pyrazine with arbitrary orientation in the first layer). Further we shall name all the essential moments as main moments ($Q_{main}$ and $d_{main}$), while the nonessential moments $Q_{xy}, Q_{xz}, Q_{yz}$ and nonessential $d$ moments as the minor ones ($Q_{minor}$ and $d_{minor}$). Since our further consideration concerns symmetrical molecules and SEHRS selection rules in these molecules, it is necessary to determine the minor and the main moments for this case. It is convenient to transfer to the values, which transform after irreducible representations of the symmetry group. Analysis of the tables of irreducible representations of all point groups [20,22] demonstrates that all the $d$ and $Q_{xy}, Q_{xz}, Q_{yz}$ moments transform after irreducible representations of the symmetry groups in the most part of the groups. Further we shall consider just such groups. The $Q_{xx}, Q_{yy}, Q_{zz}$ moments can transform after reducible representations and can be expressed via linear combinations $Q_1, Q_2, Q_3$ transforming after irreducible representations

$$Q_{xx} = a_{11}Q_1 + a_{12}Q_2 + a_{13}Q_3 \tag{24}$$

$$Q_{yy} = a_{21}Q_1 + a_{22}Q_2 + a_{23}Q_3 \tag{25}$$

$$Q_{zz} = a_{31}Q_1 + a_{32}Q_2 + a_{33}Q_3 \tag{26}$$

where the coefficients $a_{ij}$ depend on the symmetry group. The corresponding expressions for $Q_1, Q_2, Q_3$ are

$$Q_1 = b_{11}Q_{xx} + b_{12}Q_{yy} + b_{13}Q_{zz} \tag{27}$$

$$Q_2 = b_{21}Q_{xx} + b_{22}Q_{yy} + b_{23}Q_{zz} \tag{28}$$

$$Q_3 = b_{31}Q_{xx} + b_{32}Q_{yy} + b_{33}Q_{zz} \tag{29}$$

Here the coefficients $b_{ij}$ depend on the symmetry group too. The specific form of $Q_1, Q_2, Q_3$ for some point groups one can find in [20,22]. There are combinations with a constant sign, which we can name as the main moments and of a changeable sign, which can be named as the minor ones. In accordance with our previous consideration the main moments are responsible for the the strong enhancement, while the minor moments are nonessential for the scattering. For trans-1,2-bis (4-pyridyle) ethylene, pyridine and pyrazine molecules, which are considered in this paper, $Q_1, Q_2$ and $Q_3$ coincide with $Q_{xx}, Q_{yy}$ and $Q_{zz}$.

## 6. ANALYTICAL EXPRESSIONS FOR THE WAVEFUNCTIONS OF SYMMETRICAL MOLECULES

In order to obtain the expression for the SER cross-section it is necessary to know the molecular wavefunctions that take into account the vibrations of nuclei. They can be obtained in the framework of adiabatic perturbation theory [23, 24] by solving of Shrodinger equation for the molecule

$$\widehat{H}_{mol} \Psi = E \Psi \ . \tag{30}$$

Here

$$\widehat{H}_{mol} = \widehat{H}_e + \widehat{H}_n + \widehat{H}_{e-n} \ . \tag{31}$$

$$\widehat{H}_e = -\hbar^2/2m \sum_i \Delta_{\bar{r}_i} + \frac{1}{2} \sum_{\substack{i,k \\ i \neq k}} \frac{e^2}{r_{ik}} - \sum_{iJ} \frac{e^2 Z_J^*}{\left|\overline{R}_{iJ}^0\right|} \ . \tag{32}$$

$$\widehat{H}_n = -\frac{\hbar^2}{2} \sum_J \frac{1}{M_J} \Delta_{\bar{R}_J} + \frac{1}{2} \sum_{\substack{J,K \\ J \neq K}} \frac{e^2 Z_J^* Z_K^*}{\left|\overline{R}_{JK}\right|} \ , \tag{33}$$

$$\widehat{H}_{e-n} = -\sum_{iJ} \frac{e^2 Z_J^*}{\left|\overline{R}_{iJ}\right|} + \sum_{iJ} \frac{e^2 Z_J^*}{\left|\overline{R}_{iJ}^0\right|} \ , \tag{34}$$

$\widehat{H}_e, \widehat{H}_n, \widehat{H}_{e-n}$ - are the Hamiltonians of electrons in the field of the motionless nuclei, the nuclei and the Hamiltonian of interaction of electrons with nuclei respectively. $\bar{r}_i$ is a radius vector of the $i$ electron, $r_{ik}$ - is the distance between $i$ and $k$ electrons, $\overline{R}_{iJ}^o$ is the radius vector between the $i$ electron and $J$ motionless nucleus. $\overline{R}_J$ - is the radius vector of the $J$ nucleus, $\overline{R}_{JK}$ is the radius vector between the $J$ and $K$ nuclei, $M_J$ - is the mass of the $J$ nucleus, $Z_J^*$ is the atomic number. All other designations are conventional. In accordance with adiabatic perturbation theory $\hat{H}_e$ in (31) can be considered as zero approximation, and $\hat{H}_n + \hat{H}_{e-n}$ as a perturbation. The second term in (33) can be expanded by direct expansion of $1/|\overline{R}_{JK}|$ in the Taylor series. The $1/|\overline{R}_{iJ}|$ cannot be expanded directly, because there is a region of radius vectors $\overline{r}_i$ where the inequality

$$\left|\overline{R}_J^0 - \overline{r}_i\right| > \left|\Delta\overline{R}_J\right| \tag{35}$$

is not valid.

Therefore let us use the following method to solve (30). Let us try to find the solution in the form

$$\Psi = a^{(0)}\Psi^{(0)} + \Delta\Psi \tag{36}$$

with the following expression used for the energy

$$E = E^{(0)} + \Delta E \tag{37}$$

Here $\Psi^{(0)}$ satisfies the equation

$$\hat{H}_e \Psi^{(0)} = E^{(0)}\Psi^{(0)} \tag{38}$$

$a^{(0)}$ is some coefficient, $\Delta\Psi$ is an additional part of the molecule wave function, arising due to vibrations, $E^{(0)}$ is the energy of unperturbed Hamiltonian, $\Delta E$ is perturbation of the energy. Substituting (36) and (37) in (30) we can obtain, taking into account (31)

$$(\hat{H}_e - E^{(0)})\Delta\Psi = -(\hat{H}_n + \hat{H}_{e-n} - \Delta E)(a^{(0)}\Psi^0 + \Delta\Psi) \tag{39}$$

Let us designate the eigenfunctions of (38) as $\Psi_n^{(0)}$ and eigenvalues as $E_n^{(0)}$, The additional part $\Delta\Psi_n$ can be expanded in terms of the eigenfunctions of equation (38)

$$\Delta\Psi_n = \sum_{\substack{m \\ m \neq n}} a_{nm} \Psi_m^{(0)} \tag{40}$$

Substituting (40) in (39) and projecting on $\Psi_n^{(0)}$ we can obtain the following equation.

$$(\hat{H}_n + \langle n|\hat{H}_{e-n}|n\rangle - \Delta E_n) a_n^{(0)} + \sum_{\substack{m \\ m \neq n}} \langle n|\hat{H}_{e-n}|m\rangle a_{nm} = 0 \tag{41}$$

For the projection on $\Psi_l^{(0)}$ ($l \neq n$) we can obtain

$$(E_l^{(0)} - E_n^{(0)}) a_{nl} = -\langle l|\hat{H}_{e-n}|n\rangle a_n^{(0)} - \sum_{\substack{m \\ m \neq n}} \langle l|\hat{H}_n + \hat{H}_{e-n}|m\rangle a_{nm} + \Delta E_n a_{nl}$$

$$\tag{42}$$

Let us pass to another variables

$$\Delta \overline{R}_J = \lambda' \overline{U}_J \tag{43}$$

where

$$\lambda' = \sqrt[4]{\frac{m}{\overline{M}}} \tag{44}$$

is a new parameter of the adiabatic perturbation theory and $\overline{M}$ is a mean mass of nuclei. Equations (41) and (42) can be expanded in powers of this small parameter. Therefore it is necessary to introduce the following expansions for $a_{nl}$ and $\Delta E_n$

$$a_{nl} = \sum_{k=1}^{\infty} a_{nl}^{(k)} \lambda'^k \tag{45}$$

$$\Delta E_n = \sum_{k=1}^{\infty} E_n^{(k)} \lambda'^k \tag{46}$$

The matrix element $\langle n|\hat{H}_n|n\rangle = \hat{H}_n$ can be expanded in an analytical expression, which can be obtained expanding the functions $1/|\overline{R}_{JK}|$

$$\hat{H}_n = 1/2 \sum_{\substack{JK \\ J \neq K}} \frac{Z_J^* Z_K^* e^2}{\left|\overline{R}_J^0 - \overline{R}_K^0\right|} -$$

$$-1/2 \lambda' \sum_{\substack{JK \\ J \neq K}} \frac{e^2 Z_J^* Z_K^* ((\overline{R}_J^0 - \overline{R}_K^0)(\Delta \overline{U}_J - \Delta \overline{U}_K))}{\left|\overline{R}_J^0 - \overline{R}_K^0\right|^3} +$$

$$\lambda'^2 \begin{bmatrix} -\frac{\hbar^2}{2m} \sum_J \frac{\overline{M}}{M_J} \Delta_{\overline{U}_J} - 1/4 \sum_{JK} \frac{e^2 Z_J^* Z_K^* \left|\Delta \overline{U}_J - \Delta \overline{U}_K\right|^2}{\left|\overline{R}_J^0 - \overline{R}_K^0\right|^3} + \\ 3/2 \sum_{\substack{JK \\ J \neq K}} \frac{e^2 Z^* Z^* ((\overline{R}_J^0 - \overline{R}_K^0)(\Delta \overline{U}_J - \Delta \overline{U}_K))^2}{\left|\overline{R}_J^0 - \overline{R}_K^0\right|^5} \end{bmatrix} \qquad (47)$$

From the expression (34) we can see, that the expansion of matrix elements $\langle l | \hat{H}_{e-n} | n \rangle$ begins from the first power of $\Delta \overline{R}$ and hence from the first power of $\lambda'$. Then

$$\langle l | \hat{H}_{e-n} | n \rangle = \lambda' \sum_{J\alpha} \frac{\partial \langle l | \hat{H}_{e-n} | n \rangle}{\partial X_{J\alpha}} \Delta U_{J\alpha} + \\ + \lambda'^2 \sum_{\substack{JK \\ \alpha\beta}} 1/2 \frac{\partial \langle l | \hat{H}_{e-n} | n \rangle}{\partial X_{J\alpha} \partial X_{K\beta}} \Delta U_{J\alpha} \Delta U_{K\beta} \cdots \cdots \qquad (48)$$

The terms which are linear in deviations, can be approximated in the following manner. Let us expand formally the terms $1/\left|\overline{R}_{iJ}\right|$ in expression (34) in powers of $\Delta X_{J\alpha}$ and keep only the terms, which are linear in $\Delta X_{J\alpha}$. Then we can write an approximate relation

$$\frac{\partial \langle l | \hat{H}_{e-n} | n \rangle}{\partial X_{J\alpha}} \cong \sum_i \int \frac{e^2 Z_J^*}{\left|\overline{R}_{iJ}^0\right|^3} (X_{J\alpha}^0 - x_{i\alpha}) \Psi_l^* \Psi_n \times d^3 r_1 .. d^3 r_N \qquad (49)$$

The following condition should be noted, because the nuclei are in equilibrium positions

$$-\sum_K \frac{e^2 Z_J^* Z_K^* (X_{J\alpha}^0 - X_{K\alpha}^0)}{\left|\overline{R}_J^0 - \overline{R}_K^0\right|^3} +$$

$$+\sum_i \int \frac{|\Psi_n|^2 e^2 Z_J^* (X_{J\alpha}^0 - x_{i\alpha})}{\left|\overline{R}_{iJ}^0\right|^3} d^3 r_1 \ldots d^3 r_N = 0 \tag{50}$$

Using expansions (45 – 48), substituting them in (41) and (42) we can obtain relations for determination of $a_n^{(0)}$, $a_{nl}^{(k)}$ and $\mathrm{E}_n^{(k)}$. From expansion (41) for the first power of $\lambda'$ we can obtain the expression for the energy

$$\mathrm{E}_n^{(1)} = 0 \tag{51}$$

and from (42) the coefficients

$$a_{nl}^{(1)} = \frac{1}{(\mathrm{E}_n^{(0)} - \mathrm{E}_l^{(0)})} \sum_{J\alpha} \frac{\partial \langle l|\hat{\mathrm{H}}_{e-n}|n\rangle}{\partial X_{J\alpha}} \Delta U_{J\alpha} a_n^{(0)} \tag{52}$$

From expansion (41) for the second power of $\lambda'$ we can obtain an equation for determination of $a_n^{(0)}$

$$\left[ -\frac{\hbar^2}{2} \sum_J \frac{1}{M_j} \Delta_{R_J} - 1/4 \sum_{\substack{JK \\ J \neq K}} \frac{e^2 Z_J^* Z_K^* \left|\Delta \overline{R}_J - \Delta \overline{R}_K\right|^2}{\left|\overline{R}_J^0 - \overline{R}_K^0\right|^3} + \right.$$

$$+ \frac{3}{2} \sum_{\substack{JK \\ J \neq K}} \frac{e^2 Z_J^* Z_K^* ((\overline{R}_J^0 - \overline{R}_K^0)(\Delta \overline{R}_J - \Delta \overline{R}_K))^2}{\left|\overline{R}_J^0 - \overline{R}_K^0\right|^5} +$$

$$+ \sum_{\substack{l \\ l \neq n}} \sum_{\substack{JK \\ \alpha\beta}} \frac{1}{(\mathrm{E}_n^{(0)} - \mathrm{E}_l^{(0)})} \frac{\partial \langle l|\hat{\mathrm{H}}_{e-n}|n\rangle}{\partial X_{J\alpha}} \times \frac{\partial \langle n|\hat{\mathrm{H}}_{e-n}|l\rangle}{\partial X_{K\beta}} \Delta X_{J\alpha} \Delta X_{K\beta} +$$

$$+\sum_{\substack{JK \\ \alpha\beta}} \frac{1}{2} \frac{\partial \langle n|\hat{H}_{e-n}|n\rangle}{\partial X_{J\alpha} \partial X_{K\beta}} \Delta X_{J\alpha} \Delta X_{K\beta} \Big] a_n^{(0)} = \lambda'^2 E_n^{(2)} a_n^{(0)} \qquad (53)$$

All the terms, except of the first in (53) are parts of the potential. Equation (53) gives a solution for $a_n^{(0)}$ as a function of deviations of nuclei from their equilibrium positions. The corresponding equation in normal coordinates $\xi_s$ with $s$ normal vibration is the following

$$\left[ -\frac{\hbar^2}{2} \sum_s \frac{\partial^2}{\partial \xi_s^2} + \sum_s \frac{\omega_s^2 \xi_s^2}{2} \right] a_n^{(0)} = \lambda'^2 E_n^{(2)} a_n^{(0)} \qquad (54)$$

with the solution

$$a_n^{(0)} = \alpha_{\overline{V}} = \prod_s N_s H_{V_s}\left(\sqrt{\frac{\omega_s}{\hbar}} \xi_s\right) \exp\left(-\frac{\omega_s \xi_s^2}{2\hbar}\right) \qquad (55)$$

Here $V_s$ is a quantum number of a normal vibration. The set of $V_s$ can be designated as a vector $\overline{V}$. $H_{V_s}$ is a Hermitian polynomial of the $V_s$ power, $N_s$ is a normalization constant. The energy of vibrations

$$E_{\overline{V}} = \lambda'^2 E_n^{(2)} = \sum_s (V_s + 1/2)\hbar\omega_s \qquad (56)$$

The values of deviations are determined by the following formulae.

$$\Delta \overline{R}_J = \sum_s \xi_s \overline{R}_{Js}$$
$$\Delta X_{J\alpha} = \sum_s \xi_s \overline{X}_{Js\alpha} \qquad (57)$$

$\overline{R}_{Js}$ and $\overline{X}_{Js\alpha}$ are the displacement vector and its $\alpha$ projection of the $J$ nucleus in the $s$ vibration mode. Substituting (57) in (52) one can rewrite the expressions for the coefficients $a_{nl}^{(1)}$ in the following form

$$\lambda' a_{nl}^{(1)} = \frac{\sum_s \sqrt{\frac{\omega_s}{\hbar}} \xi_s R_{nls}}{(E_n^{(0)} - E_l^{(0)})} a_n^{(0)} \tag{58}$$

where

$$R_{nls} = \sum_{J\alpha} \sqrt{\frac{\hbar}{\omega_s}} \overline{X}_{Js\alpha} \frac{\partial \langle l | \hat{H}_{e-n} | n \rangle}{\partial X_{J\alpha}} \quad . \tag{59}$$

Then the general form of the wave function in the first approximation can be written as

$$\Psi_{n\overline{V}} = \left[ \Psi_n^{(0)} + \sum_{\substack{l \\ l \neq n}} \frac{\sum_s R_{nls} \sqrt{\frac{\omega_s}{\hbar}} \xi_s \Psi_l^{(0)}}{(E_n^{(0)} - E_l^{(0)})} \right] \alpha_{\overline{V}} \exp-(iE_{n\overline{V}}t)/\hbar \tag{60}$$

where

$$E_{n\overline{V}} = E_n^{(0)} + \sum_s \hbar \omega_s (V_s + 1/2) \tag{61}$$

The term of the Hamiltonian (34) describes interaction of electrons with molecular vibrations or molecular phonons. Owing to the vibrations the potential of the nuclei changes by the value of deformation potential. Therefore we can consider the expansion of matrix elements $\langle l | \hat{H}_{e-n} | n \rangle$ in (48) as matrix elements of deformation potential, which cause electron transitions in the electron shell. The addition to the wave function in (60) is due to deformation potential in a linear approximation. In the zero approximation the ground state of the molecule is described by the electron function $\Psi_n^{(0)}$ with the energy $E_n^{(0)}$ and the wave function of nuclei $\alpha_{\overline{V}}$ with the energy (56). Owing to the deformation potential the electron shell can absorb or emit a molecular phonon (M. ph.). These processes arise due to the second term in (60), when the only nonzero terms in matrix elements of vibrational quantum transitions are those with the change of one of the vibrational quantum numbers on one unit.

To calculate the SER cross-section, we need expressions for the virtual electronic and vibrational states. Their expressions can be obtained as those for the wave function of the ground state and can be written by changing of indices in corresponding expressions.

$$\Psi_{m\bar{V}} = \left[ \Psi_m^{(0)} + \sum_{\substack{k \\ k \neq m}} \frac{\sum_s R_{mks} \sqrt{\frac{\omega_s}{\hbar}} \xi_s \Psi_k^{(0)}}{(E_m^{(0)} - E_k^{(0)})} \right] \alpha_{\bar{V}} \exp{-(iE_{m\bar{V}}t)/\hbar} \qquad (62)$$

Here we neglect by changes of the frequencies and vibrational wave functions for the virtual states compared with the ground state. The expression for the coefficients $R_{mks}$ is the following

$$R_{mks} = \sum_{J\alpha} \sqrt{\frac{\hbar}{\omega_s}} \overline{X}_{Js\alpha} \frac{\partial \langle k | \hat{H}_{e-n} | m \rangle}{\partial X_{J\alpha}} \qquad (63)$$

The expressions for the wavefunctions of a symmetrical molecule can be obtained from (60) by changing of the index $s$ by double indices $s$ and $p$. Here index $s$ numerates the groups of degenerated states, while the index $p$ the states inside the group. Then the corresponding expressions for the wavefunctions of the ground and excided states have the following form

$$\Psi_{n\bar{V}} = \left[ \Psi_n^{(0)} + \sum_{\substack{l \\ l \neq n}} \frac{\sum_{s,p} R_{nl(s,p)} \sqrt{\frac{\omega_{(s,p)}}{\hbar}} \xi_{(s,p)} \Psi_l^{(0)}}{(E_n^{(0)} - E_l^{(0)})} \right] \alpha_{\bar{V}} \exp{-(iE_{n\bar{V}}t)/\hbar} \qquad (64)$$

$$\Psi_{m\bar{V}} = \left[ \Psi_m^{(0)} + \sum_{\substack{k \\ k \neq m}} \frac{\sum_{s,p} R_{mk(s,p)} \sqrt{\frac{\omega_{(s,p)}}{\hbar}} \xi_{(s,p)} \Psi_k^{(0)}}{(E_m^{(0)} - E_k^{(0)})} \right] \alpha_{\bar{V}} \exp(-iE_{m\bar{V}}t)/\hbar \qquad (65)$$

The corresponding form for the values $R_{nl(s,p)}, \alpha_{\bar{V}}$ and $E_{m\bar{V}}$ is

$$E_{m,\bar{V}} = E_m^{(0)} + \sum_{s,p} \hbar \omega_{(s,p)} (V_{(s,p)} + 1/2) \qquad (66)$$

$$\alpha_{\overline{V}} = \prod_{(s,p)} N_{(s,p)} H_{V_{(s,p)}} \left( \sqrt{\frac{\omega_{(s,p)}}{\hbar}} \xi_{(s,p)} \right) exp\left( -\frac{\omega_{(s,p)} \xi_{(s,p)}^2}{2\hbar} \right) \quad (67)$$

$$R_{mk(s,p)} = \sum_{J\alpha} \sqrt{\frac{\hbar}{\omega_{(s,p)}}} \overline{X}_{J(s,p)\alpha} \frac{\partial \langle k|\hat{H}_{e-n}|m\rangle}{\partial X_{J\alpha}} \quad (68)$$

## 7. THE EXPRESSION FOR THE SEHRS CROSS-SECTION

The expression for the SEHR cross-section of symmetrical molecules can be obtained in the same manner as the cross-section for SERS [10,13]. Since SEHRS is a three quantum process the cross-section can be obtained using time dependent quantum mechanical perturbation theory. It is expressed via the

$$\frac{d}{dt} \left| w^{(3)}_{(n,\overline{V}\pm 1),(n,\overline{V})}(t) \right|^2, \quad (69)$$

where $w^{(3)}_{(n,\overline{V}\pm 1),(n,\overline{V})}(t)$ is the third term in expansion of perturbation coefficient between the states with $(n, \overline{V} \pm 1)$ and $(n, \overline{V})$ quantum numbers, which refer to the Stokes and AntiStokes scattering respectively. Here the expression $(\overline{V} \pm 1)$ means the change of one of the vibrational quantum numbers on one unit.

$$w^{(3)}_{(n,\overline{V}\pm 1),(n,\overline{V})}(t) = \left(-\frac{i}{\hbar}\right)^3 \sum_{\substack{m \\ m \neq n}} \left[ \int_0^t \langle n, \overline{V} \pm 1 | \hat{H}^{inc}_{e-r} + \hat{H}^{scat}_{e-r} | k, \overline{V} \pm 1 \rangle dt_1 \times \right.$$

$$\times \int_0^{t_1} \langle k, \overline{V} \pm 1 | \hat{H}^{inc}_{e-r} + \hat{H}^{scat}_{e-r} | m, \overline{V} \pm 1 \rangle dt_2 \times \int_0^{t_2} \langle m, \overline{V} \pm 1 | \hat{H}^{inc}_{e-r} + \hat{H}^{scat}_{e-r} | n, \overline{V} \rangle dt_3 +$$

$$+ \int_0^t \langle n, \overline{V} \pm 1 | \hat{H}^{inc}_{e-r} + \hat{H}^{scat}_{e-r} | k, \overline{V} \pm 1 \rangle dt_1 \times$$

$$\times \int_0^{t_1} \langle k, \overline{V} \pm 1 | \hat{H}_{e-r}^{inc} + \hat{H}_{e-r}^{scat} | m, \overline{V} \rangle dt_2 \times \int_0^{t_2} \langle m, \overline{V} | \hat{H}_{e-r}^{inc} + \hat{H}_{e-r}^{scat} | n, \overline{V} \rangle dt_3 +$$

$$+ \int_0^{t} \langle n, \overline{V} \pm 1 | \hat{H}_{e-r}^{inc} + \hat{H}_{e-r}^{scat} | k, \overline{V} \rangle dt_1 \times$$

$$\times \int_0^{t_1} \langle k, \overline{V} | \hat{H}_{e-r}^{inc} + \hat{H}_{e-r}^{scat} | m, \overline{V} \rangle dt_2 \times \int_0^{t_2} \langle m, \overline{V} | \hat{H}_{e-r}^{inc} + \hat{H}_{e-r}^{scat} | n, \overline{V} \rangle dt_3 \Bigg] . \tag{70}$$

Using the expressions (14, 15, 64-67) one can find the expression for the SEHRS cross-section

$$d\sigma_{SEHRS}\begin{pmatrix} St \\ AnSt \end{pmatrix} = \frac{\omega_{inc}\omega_{scat}^3}{64\pi^2 \hbar^4 \varepsilon_0^2 c^4} \left| \overline{E}_{0,inc} \right|_{vol}^2 \frac{\left| \overline{E}_{inc} \right|_{surf}^2 \left| \overline{E}_{inc} \right|_{surf}^2 \left| \overline{E}_{scat} \right|_{surf}^2}{\left| \overline{E}_{0,inc} \right|_{vol}^2 \left| \overline{E}_{0,inc} \right|_{vol}^2 \left| \overline{E}_{0,scat} \right|_{vol}^2} \times$$

$$\times \begin{pmatrix} \frac{V_{(s,p)}+1}{2} \\ \frac{V_{(s,p)}}{2} \end{pmatrix} \times \left| C_{V_{(s,p)}}[(\overline{e}\overline{f}_e)_{inc},(\overline{e}\overline{f}_e)_{inc},(\overline{e}^*\overline{f}_e^*)_{scat}]\begin{pmatrix} St \\ anSt \end{pmatrix} \right|^2 dO.$$

$$\tag{71}$$

Here the signs *surf* and *vol* designate that the corresponding fields refer to the surface field and the field in a free space. $\left(\overline{E}_{inc}\right)_{surf}$ - is the surface field, which arise due to the incident field $\left(\overline{E}_{inc}\right)_{vol}$, $\left(\overline{E}_{scat}\right)_{surf}$ - is the surface field, which arise due to the incident field $\left(\overline{E}_{scat}\right)_{vol}$, from the direction, when the direction of the wave reflected from the surface coincides with the direction of the scattering. $O$ - is a solid angle. Here

$$C_{V_{(s,p)}}[f_1, f_2, f_3]\begin{pmatrix} St \\ AnSt \end{pmatrix} = \sum_{m,r,l \neq n} \frac{\langle n|f_3|m\rangle\langle m|f_2|r\rangle\langle r|f_1|l\rangle R_{n,l,(s,p)}}{(E_n^{(0)} - E_l^{(0)})(\omega_{m,n} - 2\omega_{inc})(\omega_{r,n} \pm \omega_{(s,p)} - \omega_{inc})}$$

$$+ \sum_{m,r,l \neq n} \frac{\langle n|f_2|m\rangle\langle m|f_3|r\rangle\langle r|f_1|l\rangle R_{n,l,(s,p)}}{(E_n^{(0)} - E_l^{(0)})(\omega_{m,n} + \omega_{scat} - \omega_{inc})(\omega_{r,n} \pm \omega_{(s,p)} - \omega_{inc})} +$$

$$+ \sum_{m,r,l \neq n} \frac{\langle n|f_2|m\rangle\langle m|f_1|r\rangle\langle r|f_3|l\rangle R_{n,l,(s,p)}}{(\mathrm{E}_n^{(0)} - \mathrm{E}_l^{(0)})(\omega_{m,n} + \omega_{scat} - \omega_{inc})(\omega_{r,n} \pm \omega_{(s,p)} + \omega_{scat})} +$$

$$+ \sum_{m,r,l \neq n} \frac{\langle n|f_3|m\rangle\langle m|f_2|r\rangle R^*_{r,l,(s,p)}\langle l|f_1|n\rangle}{(\mathrm{E}_r^{(0)} - \mathrm{E}_l^{(0)})(\omega_{m,n} - 2\omega_{inc})(\omega_{r,n} \mp \omega_{(s,p)} - \omega_{inc})} +$$

$$+ \sum_{m,r,l \neq n} \frac{\langle n|f_2|m\rangle\langle m|f_3|r\rangle R^*_{r,l,(s,p)}\langle l|f_1|n\rangle}{(\mathrm{E}_r^{(0)} - \mathrm{E}_{(l}^{(0)})(\omega_{m,n} + \omega_{scat} - \omega_{inc})(\omega_{r,n} \mp \omega_{(s,p)} - \omega_{inc})} +$$

$$+ \sum_{m,r,l \neq n} \frac{\langle n|f_2|m\rangle\langle m|f_1|r\rangle R^*_{r,l,(s,p)}\langle l|f_3|n\rangle}{(\mathrm{E}_r^{(0)} - \mathrm{E}_l^{(0)})(\omega_{m,n} + \omega_{scat} - \omega_{inc})(\omega_{r,n} \mp \omega_{(s,p)} + \omega_{scat})} +$$

$$+ \sum_{m,r,l \neq n} \frac{\langle n|f_3|m\rangle\langle m|f_2|l\rangle R_{r,l,(s,p)}\langle r|f_1|n\rangle}{(\mathrm{E}_r^{(0)} - \mathrm{E}_l^{(0)})(\omega_{m,n} \pm \omega_{(s,p)} - 2\omega_{inc})(\omega_{r,n} - \omega_{inc})} +$$

$$+ \sum_{m,r,l \neq n} \frac{\langle n|f_2|m\rangle\langle m|f_3|l\rangle R_{r,l,(s,p)}\langle r|f_1|n\rangle}{(\mathrm{E}_r^{(0)} - \mathrm{E}_l^{(0)})(\omega_{m,n} + \omega_{scat} \pm \omega_{(s,p)} - \omega_{inc})(\omega_{r,n} - \omega_{inc})} +$$

$$+ \sum_{m,r,l \neq n} \frac{\langle n|f_2|m\rangle\langle m|f_1|l\rangle R_{r,l,(s,p)}\langle r|f_3|n\rangle}{(\mathrm{E}_r^{(0)} - \mathrm{E}_l^{(0)})(\omega_{m,n} + \omega_{scat} \pm \omega_{(s,p)} - \omega_{inc})(\omega_{r,n} + \omega_{scat})} +$$

$$+ \sum_{m,r,l \neq n} \frac{\langle n|f_3|m\rangle R^*_{m,l,(s,p)}\langle l|f_2|r\rangle\langle r|f_1|n\rangle}{(\mathrm{E}_m^{(0)} - \mathrm{E}_l^{(0)})(\omega_{m,n} \mp \omega_{(s,p)} - 2\omega_{inc})(\omega_{r,n} - \omega_{inc})} +$$

$$+ \sum_{m,r,l \neq n} \frac{\langle n|f_2|m\rangle R^*_{m,l,(s,p)}\langle l|f_3|r\rangle\langle r|f_1|n\rangle}{(\mathrm{E}_m^0 - \mathrm{E}_l^0)(\omega_{m,n} + \omega_{scat} \mp \omega_{(s,p)} - \omega_{inc})(\omega_{r,n} - \omega_{inc})} +$$

$$+ \sum_{m,r,l \neq n} \frac{\langle n|f_2|m\rangle R^*_{m,l,(s,p)}\langle l|f_1|r\rangle\langle r|f_3|n\rangle}{(\mathrm{E}_m^{(0)} - \mathrm{E}_l^{(0)})(\omega_{m,n} + \omega_{scat} \mp \omega_{(s,p)} - \omega_{inc})(\omega_{r,n} + \omega_{scat})} +$$

$$+ \sum_{m,r,l \neq n} \frac{\langle n|f_3|l\rangle R_{m,l,(s,p)}\langle m|f_2|r\rangle\langle r|f_1|n\rangle}{(\mathrm{E}_m^{(0)} - \mathrm{E}_l^{(0)})(\omega_{m,n} - 2\omega_{inc})(\omega_{r,n} - \omega_{inc})} +$$

$$+ \sum_{m,r,l \neq n} \frac{\langle n|f_2|l\rangle R_{m,l,(s,p)} \langle m|f_3|r\rangle \langle r|f_1|n\rangle}{(E_m^{(0)} - E_l^{(0)})(\omega_{m,n} + \omega_{scat} - \omega_{inc})(\omega_{r,n} - \omega_{inc})} +$$

$$+ \sum_{m,r,l \neq n} \frac{\langle n|f_2|l\rangle R_{m,l,(s,p)} \langle m|f_1|r\rangle \langle r|f_1|n\rangle}{(E_m^{(0)} - E_l^{(0)})(\omega_{m,n} + \omega_{scat} - \omega_{inc})(\omega_{r,n} + \omega_{scat})} +$$

$$+ \sum_{m,r,l \neq n} \frac{R_{n,l,(s,p)}^* \langle l|f_3|m\rangle \langle m|f_2|r\rangle \langle r|f_1|n\rangle}{(E_n^{(0)} - E_l^{(0)})(\omega_{m,n} - 2\omega_{inc})(\omega_{r,n} - \omega_{inc})} +$$

$$+ \sum_{m,r,l \neq n} \frac{R_{n,l,(s,p)}^* \langle l|f_2|m\rangle \langle m|f_3|r\rangle \langle r|f_1|n\rangle}{(E_n^{(0)} - E_l^{(0)})(\omega_{m,n} + \omega_{scat} - \omega_{inc})(\omega_{r,n} - \omega_{inc})} +$$

$$+ \sum_{m,r,l \neq n} \frac{R_{n,l,(s,p)}^* \langle l|f_2|m\rangle \langle m|f_1|r\rangle \langle r|f_3|n\rangle}{(E_n^{(0)} - E_l^{(0)})(\omega_{m,n} + \omega_{scat} - \omega_{inc})(\omega_{r,n} + \omega_{scat})} \quad (72)$$

is the scattering tensor. Under $f_1, f_2, f_3$ we mean the dipole and quadrupole moments. Using the following property of the scattering tensor

$$C_{V_{(s,p)}}[f_i, f_j, (a_1 f_k + a_2 f_m)] = a_1 C_{V_{(s,p)}}[f_i, f_j, f_k] + a_2 C_{V_{(s,p)}}[f_i, f_j, f_m]$$

$$C_{V_{(s,p)}}[f_i, (a_1 f_j + a_2 f_k), f_m] = a_1 C_{V_{(s,p)}}[f_i, f_j, f_m] + a_2 C_{V_{(s,p)}}[f_i, f_k, f_m]$$

$$C_{V_{(s,p)}}[(a_1 f_i + a_2 f_j), f_k, f_m] = a_1 C_{V_{(s,p)}}[f_i, f_k, f_m] + a_2 C_{V_{(s,p)}}[f_j, f_k, f_m] \quad (73)$$

the SEHR cross-section can be written in the form

$$d\sigma_{SEHRS} \begin{Bmatrix} St \\ AnSt \end{Bmatrix} = \frac{\omega_{inc} \omega_{scat}^3}{64\pi^2 \hbar^4 \varepsilon_0^2 c^4} \left| \overline{E}_{0,inc} \right|_{vol}^2 \frac{\left| \overline{E}_{inc} \right|_{surf}^2 \left| \overline{E}_{inc} \right|_{surf}^2 \left| \overline{E}_{scat} \right|_{surf}^2}{\left| \overline{E}_{0,inc} \right|_{vol}^2 \left| \overline{E}_{0,inc} \right|_{vol}^2 \left| \overline{E}_{0,scat} \right|_{vol}^2} \times$$

$$\times \left( \frac{\frac{V_{(s,p)} + 1}{2}}{\frac{V_{(s,p)}}{2}} \right) \left| \sum_{f_1, f_2, f_3} S_{(s,p), f_1 - f_2 - f_3} \right|^2 dO,$$

(74)

Here $S_{(s,p),f_1-f_2-f_3}$ are the scattering contributions, which depend on the dipole and quadrupole moments $f_1, f_2, f_3$. Further we shall omit the indices $(s, p)$, meaning that $S$ depends on them. The sums of the contributions

$$S_{d-d-d} = \sum_{i,j,k} C_{V_{(s,p)}}[d_i, d_j, d_k] (e^*_{scat,i} e_{inc,j} e_{inc,k})_{surf} \tag{75}$$

$$S_{d-d-Q} = \sum_{i,j,\chi,\eta} C_{V_{(s,p)}}[d_i, d_j, Q_{\chi\eta}] \left( e^*_{scat,i} e_{inc,j} \frac{1}{2|\overline{E}_{inc}|} \frac{\partial E^{inc}_\chi}{\partial x_\eta} \right)_{surf} \tag{76}$$

$$S_{d-Q-d} = \sum_{i,\gamma,\delta,k} C_{V_{(s,p)}}[d_i, Q_{\gamma\delta}, d_k] \left( e^*_{scat,i} \frac{1}{2|\overline{E}_{inc}|} \frac{\partial E^{inc}_\gamma}{\partial x_\delta} e_{inc,k} \right)_{surf} \tag{77}$$

$$S_{Q-d-d} = \sum_{\alpha,\beta,j,k} C_{V_{(s,p)}}[Q_{\alpha\beta}, d_j, d_k] \left( \frac{1}{2|\overline{E}_{scat}|} \frac{\partial E^{scat^*}_\alpha}{\partial x_\beta} e_{inc,j} e_{inc,k} \right)_{surf} \tag{78}$$

$$S_{d-Q-Q} = \sum_{i,\gamma,\delta,\chi,\eta} C_{V_{(s,p)}}[d_i, Q_{\gamma\delta}, Q_{\chi\eta}] \left( e^*_{scat,i} \frac{1}{2|\overline{E}_{inc}|} \frac{\partial E^{inc}_\gamma}{\partial x_\delta} \frac{1}{2|\overline{E}_{inc}|} \frac{\partial E^{inc}_\chi}{\partial x_\eta} \right)_{surf} \tag{79}$$

$$S_{Q-d-Q} = \sum_{j,\alpha,\beta,\chi,\eta} C_{V_{(s,p)}}[Q_{\alpha\beta}, d_j, Q_{\chi\eta}] \left( \frac{1}{2|\overline{E}_{inc}|} \frac{\partial E^{scat^*}_\alpha}{\partial x_\beta} e_{inc,j} \frac{1}{2|\overline{E}_{inc}|} \frac{\partial E^{inc}_\chi}{\partial x_\eta} \right)_{surf} \tag{80}$$

$$S_{Q-Q-d} = \sum_{\alpha,\beta,\gamma,\delta,k} C_{V_{(s,p)}}[Q_{\alpha\beta}, Q_{\gamma\delta}, d_k] \left( \frac{1}{2|\overline{E}_{scat}|} \frac{\partial E^{scat^*}_\alpha}{\partial x_\beta} \frac{1}{2|\overline{E}_{inc}|} \frac{\partial E^{\delta}_\gamma}{\partial x_\delta} e_{inc,k} \right)_{surf}$$

$$\tag{81}$$

$$S_{Q-Q-Q} = \sum_{\alpha,\beta,\gamma,\delta,\chi,\eta} C_{V_{(s,p)}}[Q_{\alpha\beta}, Q_{\gamma\delta}, Q_{\chi\eta}] \left( \frac{1}{2|\overline{E}_{scat}|} \frac{\partial E^{scat^*}_\alpha}{\partial x_\beta} \frac{1}{2|\overline{E}_{inc}|} \frac{\partial E^{inc}_\gamma}{\partial x_\delta} \frac{1}{2|\overline{E}_{inc}|} \frac{\partial E^{inc}_\chi}{\partial x_\eta} \right)_{surf}$$

$$\tag{82}$$

Let us transfer to the combinations of moments, transforming after irreducible representations of the molecule symmetry group in accordance with the reasoning above and the formulae (24-26). After substitution of (24-26) in (76-82) and then in (74), the cross-section of SEHRS can be written as

$$d\sigma_{SEHRS_s}\binom{St}{AnSt} = \frac{\omega_{inc}\omega_{scat}^3}{64\pi^2\hbar^4\varepsilon_0^2 c^4}\left|\overline{E}_{inc}\right|_{vol}^2 \frac{\left|\overline{E}_{inc}\right|_{surf}^2 \left|\overline{E}_{inc}\right|_{surf}^2 \left|\overline{E}_{scat}\right|_{surf}^2}{\left|\overline{E}_{inc}\right|_{vol}^2 \left|\overline{E}_{inc}\right|_{vol}^2 \left|\overline{E}_{scat}\right|_{vol}^2} \times$$

$$\times \sum_p \left(\frac{\frac{V_{(s,p)}+1}{2}}{\frac{V_{(s,p)}}{2}}\right) \left|\sum_{f_1,f_2,f_3} T_{f_1-f_2-f_3}\binom{St}{AnSt}\right|^2 dO,$$

(83)

where

$$T_{d-d-d} = S_{d-d-d},$$ (84)

$$T_{d-d-Q} = \sum_{\substack{i,j,\chi,\eta \\ \chi\neq\eta}} C_{V_{(s,p)}}[d_i,d_j,Q_{\chi\eta}]\left(e_{scat,i}^* e_{inc,j} \frac{1}{2|\overline{E}_{inc}|} \frac{\partial E_\chi^{inc}}{\partial x_\eta}\right)_{surf} +$$

$$+ \sum_{i,j,\chi,k} a_{\chi,k} C_{V_{(s,p)}}[d_i,d_j,Q_k]\left(e_{scat,i}^* e_{inc,j} \frac{1}{2|\overline{E}_{inc}|} \frac{\partial E_\chi^{inc}}{\partial x_\chi}\right)_{surf}$$

(85)

$$T_{d-Q-d} = \sum_{\substack{i,\gamma,\delta,k \\ \gamma\neq\delta}} C_{V_{(s,p)}}[d_i,Q_{\gamma\delta},d_k]\left(e_{scat,i}^* \frac{1}{2|\overline{E}_{inc}|} \frac{\partial E_\gamma^{inc}}{\partial x_\delta} e_{inc,k}\right)_{surf} +$$

$$+ \sum_{i,\gamma,j,k} a_{\gamma,j} C_{V_{(s,p)}}[d_i,Q_j,d_k]\left(e_{scat,i}^* \frac{1}{2|\overline{E}_{inc}|} \frac{\partial E_\gamma^{inc}}{\partial x_\gamma} e_{inc,k}\right)_{surf}$$

(86)

$$T_{Q-d-d} = \sum_{\substack{\alpha,\beta,j,k \\ \alpha\neq\beta}} C_{V_{(s,p)}}[Q_{\alpha\beta},d_j,d_k]\left(\frac{1}{2|\overline{E}_{scat}|} \frac{\partial E_\alpha^{scat^*}}{\partial x_\beta} e_{inc,j} e_{inc,k}\right) +$$

$$+ \sum_{\alpha,i,j,k} a_{\alpha,i} C_{V_{(s,p)}} [Q_i, d_j, d_k] \left( \frac{1}{2|\overline{E}_{scat}|} \frac{\partial E_\alpha^{scat*}}{\partial x_\alpha} e_{inc,j} e_{inc,k} \right) \tag{87}$$

$$T_{d-Q-Q} = \sum_{\substack{i,\gamma,\delta,\chi,\eta \\ \gamma \neq \delta \\ \chi \neq \eta}} C_{V_{(s,p)}} [d_i, Q_{\gamma\delta}, Q_{\chi\eta}] \left( e_{scat,i}^* \frac{1}{2|\overline{E}_{inc}|} \frac{\partial E_\gamma^{inc}}{\partial x_\delta} \frac{1}{2|\overline{E}_{inc}|} \frac{\partial E_\chi^{inc}}{\partial x_\eta} \right)_{surf} +$$

$$+ \sum_{\substack{i,\gamma,\delta,\chi,k \\ \gamma \neq \delta}} a_{\gamma,k} C_{V_{(s,p)}} [d_i, Q_{\gamma\delta}, Q_k] \left( e_{scat,i}^* \frac{1}{2|\overline{E}_{inc}|} \frac{\partial E_\gamma^{inc}}{\partial x_\delta} \frac{1}{2|\overline{E}_{inc}|} \frac{\partial E_\chi^{inc}}{\partial x_\chi} \right)_{surf} +$$

$$+ \sum_{i,\gamma,j,\chi,\eta} a_{\gamma,j} C_{V_{(s,p)}} [d_i, Q_j, Q_{\chi\eta}] \left( e_{scat,i}^* \frac{1}{2|\overline{E}_{inc}|} \frac{\partial E_\gamma^{inc}}{\partial x_\gamma} \frac{1}{2|\overline{E}_{inc}|} \frac{\partial E_\chi^{inc}}{\partial x_\eta} \right)_{surf} +$$

$$+ \sum_{i,\gamma,j,\chi,k} a_{\gamma,j} a_{\chi,k} C_{V_{(s,p)}} [d_i, Q_j, Q_k] \left( e_{scat,i}^* \frac{1}{2|\overline{E}_{inc}|} \frac{\partial E_\gamma^{inc}}{\partial x_\gamma} \frac{1}{2|\overline{E}_{inc}|} \frac{\partial E_\chi^{inc}}{\partial x_\chi} \right)_{surf} \tag{88}$$

$$T_{Q-d-Q} = \sum_{\substack{\alpha,\beta,j,\gamma,\delta \\ \alpha \neq \beta \\ \gamma \neq \delta}} C_{V_{(s,p)}} [Q_{\alpha\beta}, d_j, Q_{\gamma\delta}] \left( \frac{1}{2|\overline{E}_{scat}|} \frac{\partial E_\alpha^{scat*}}{\partial x_\beta} e_{inc,j} \frac{1}{2|\overline{E}_{inc}|} \frac{\partial E_\gamma^{inc}}{\partial x_\delta} \right)_{surf} +$$

$$+ \sum_{\substack{\alpha,\beta,j,\gamma,k \\ \alpha \neq \beta}} a_{\gamma,k} C_{V_{(s,p)}} [Q_{\alpha\beta}, d_j, Q_k] \left( \frac{1}{2|\overline{E}_{scat}|} \frac{\partial E_\alpha^{scat*}}{\partial x_\beta} e_{inc,j} \frac{1}{2|\overline{E}_{inc}|} \frac{\partial E_\gamma^{inc}}{\partial x_\gamma} \right)_{surf} +$$

$$+ \sum_{\substack{\alpha,i,j,\chi,\eta \\ \chi \neq \eta}} a_{\alpha,i} C_{V_{(s,p)}}\left[Q_i, d_j, Q_{\chi\eta}\right]\left(\frac{1}{2|\overline{E}_{scat}|}\frac{\partial E_\alpha^{scat^*}}{\partial x_\alpha} e_{inc,j} \frac{1}{2|\overline{E}_{inc}|}\frac{\partial E_\chi^{inc}}{\partial x_\eta}\right)_{surf} +$$

$$+ \sum_{\alpha,i,j,\chi,k} a_{\alpha,i} a_{\chi,k} C_{V_{(s,p)}}\left[Q_i, d_j, Q_k\right]\left(\frac{1}{2|\overline{E}_{scat}|}\frac{\partial E_\alpha^{scat^*}}{\partial x_\alpha} e_{inc,j} \frac{1}{2|\overline{E}_{inc}|}\frac{\partial E_\chi^{inc}}{\partial x_\chi}\right)_{surf}$$

(89)

$$T_{Q-Q-d} = \sum_{\substack{\alpha,\beta,\gamma,\delta,k \\ \alpha \neq \beta \\ \gamma \neq \delta}} C_{V_{(s,p)}}\left[Q_{\alpha\beta}, Q_{\gamma\delta}, d_k\right]\left(\frac{1}{2|\overline{E}_{scat}|}\frac{\partial E_\alpha^{scat^*}}{\partial x_\beta} \frac{1}{2|\overline{E}_{inc}|}\frac{\partial E_\gamma^{inc}}{\partial x_\delta} e_{inc,k}\right)_{surf} +$$

$$+ \sum_{\substack{\alpha,\beta,\gamma,j,k \\ \alpha \neq \beta}} a_{\gamma,j} C_{V_{(s,p)}}\left[Q_{\alpha\beta}, Q_j, d_k\right]\left(\frac{1}{2|\overline{E}_{scat}|}\frac{\partial E_\alpha^{scat^*}}{\partial x_\beta} \frac{1}{2|\overline{E}_{inc}|}\frac{\partial E_\gamma^{inc}}{\partial x_\gamma} e_{inc,k}\right)_{surf} +$$

$$+ \sum_{\substack{\alpha,i,\gamma,\delta,k \\ \gamma \neq \delta}} a_{\alpha,i} C_{V_{(s,p)}}\left[Q_i, Q_{\gamma\delta}, d_k\right]\left(\frac{1}{2|\overline{E}_{scat}|}\frac{\partial E_\alpha^{scat^*}}{\partial x_\alpha} \frac{1}{2|\overline{E}_{inc}|}\frac{\partial E_\gamma^{inc}}{\partial x_\delta} e_{inc,k}\right)_{surf} +$$

$$+ \sum_{\alpha,i,\gamma,j,k} a_{\alpha,i} a_{\gamma,j} C_{V_{(s,p)}}\left[Q_i, Q_j, d_k\right]\left(\frac{1}{2|\overline{E}_{scat}|}\frac{\partial E_\alpha^{scat^*}}{\partial x_\alpha} \frac{1}{2|\overline{E}_{inc}|}\frac{\partial E_\gamma^{inc}}{\partial x_\gamma} e_{inc,k}\right)_{surf}$$

(90)

$$T_{Q-Q-Q} = \sum_{\substack{\alpha,\beta,\gamma,\delta,\chi,\eta \\ \alpha \neq \beta \\ \gamma \neq \delta \\ \chi \neq \eta}} C_{V_{(s,p)}} [Q_{\alpha\beta}, Q_{\gamma\delta}, Q_{\chi\eta}] \left( \frac{1}{2|\overline{E}_{scat}|} \frac{\partial E_\alpha^{scat^*}}{\partial x_\beta} \frac{1}{2|\overline{E}_{inc}|} \frac{\partial E_\gamma^{inc}}{\partial x_\delta} \frac{1}{2|\overline{E}_{inc}|} \frac{\partial E_\chi^{inc}}{\partial x_\eta} \right)_{surf} +$$

$$+ \sum_{\substack{\alpha,\beta,\gamma,\delta,\chi,k \\ \alpha \neq \beta \\ \gamma \neq \delta}} a_{\chi,k} C_{V_{(s,p)}} [Q_{\alpha\beta}, Q_{\gamma\delta}, Q_k] \left( \frac{1}{2|\overline{E}_{scat}|} \frac{\partial E_\alpha^{scat^*}}{\partial x_\beta} \frac{1}{2|\overline{E}_{inc}|} \frac{\partial E_\gamma^{inc}}{\partial x_\delta} \frac{1}{2|\overline{E}_{inc}|} \frac{\partial E_\chi^{inc}}{\partial x_\chi} \right)_{surf} +$$

$$+ \sum_{\substack{\alpha,\beta,\gamma,j,\chi,\eta \\ \alpha \neq \beta \\ \chi \neq \eta}} a_{\gamma,j} C_{V_{(s,p)}} [Q_{\alpha\beta}, Q_j, Q_{\chi\eta}] \left( \frac{1}{2|\overline{E}_{scat}|} \frac{\partial E_\alpha^{scat^*}}{\partial x_\beta} \frac{1}{2|\overline{E}_{inc}|} \frac{\partial E_\gamma^{inc}}{\partial x_\gamma} \frac{1}{2|\overline{E}_{inc}|} \frac{\partial E_\chi^{inc}}{\partial x_\eta} \right)_{surf} +$$

$$+ \sum_{\substack{\alpha,i,\gamma,\delta,\chi,\eta \\ \gamma \neq \delta \\ \chi \neq \eta}} a_{\alpha,i} C_{V_{(s,p)}} [Q_i, Q_{\gamma\delta}, Q_{\chi\eta}] \left( \frac{1}{2|\overline{E}_{scat}|} \frac{\partial E_\alpha^{scat^*}}{\partial x_\alpha} \frac{1}{2|\overline{E}_{inc}|} \frac{\partial E_\gamma^{inc}}{\partial x_\delta} \frac{1}{2|\overline{E}_{inc}|} \frac{\partial E_\chi^{inc}}{\partial x_\eta} \right)_{surf} +$$

$$+ \sum_{\substack{\alpha,\beta,\gamma,j,\chi,k \\ \alpha \neq \beta}} a_{\gamma,j} a_{\chi,k} C_{V_{(s,p)}} [Q_{\alpha\beta}, Q_j, Q_k] \left( \frac{1}{2|\overline{E}_{scat}|} \frac{\partial E_\alpha^{scat^*}}{\partial x_\beta} \frac{1}{2|\overline{E}_{inc}|} \frac{\partial E_\gamma^{inc}}{\partial x_\gamma} \frac{1}{2|\overline{E}_{inc}|} \frac{\partial E_\chi^{inc}}{\partial x_\chi} \right)_{surf} +$$

$$+ \sum_{\substack{\alpha,i,\gamma,\delta,\chi,k \\ \gamma \neq \delta}} a_{\alpha,i} a_{\chi,k} C_{V_{(s,p)}} [Q_i, Q_{\gamma\delta}, Q_k] \left( \frac{1}{2|\overline{E}_{scat}|} \frac{\partial E_\alpha^{scat^*}}{\partial x_\alpha} \frac{1}{2|\overline{E}_{inc}|} \frac{\partial E_\gamma^{inc}}{\partial x_\delta} \frac{1}{2|\overline{E}_{inc}|} \frac{\partial E_\chi^{inc}}{\partial x_\chi} \right)_{surf} +$$

$$+ \sum_{\substack{\alpha,i,\gamma,j,\chi,\eta \\ \chi \neq \eta}} a_{\alpha,i} a_{\gamma,j} C_{V_{(s,p)}} [Q_i, Q_j, Q_{\chi\eta}] \left( \frac{1}{2|\overline{E}_{scat}|} \frac{\partial E_\alpha^{scat^*}}{\partial x_\alpha} \frac{1}{2|\overline{E}_{inc}|} \frac{\partial E_\gamma^{inc}}{\partial x_\gamma} \frac{1}{2|\overline{E}_{inc}|} \frac{\partial E_\chi^{inc}}{\partial x_\eta} \right)_{surf} +$$

$$+ \sum_{\alpha,i,\gamma,j,\chi,k} a_{\alpha,i} a_{\gamma,j} a_{\chi,k} C_{V_{(s,p)}} [Q_i, Q_j, Q_k] \left( \frac{1}{2|\overline{E}_{scat}|} \frac{\partial E_\alpha^{scat^*}}{\partial x_\alpha} \frac{1}{2|\overline{E}_{inc}|} \frac{\partial E_\gamma^{inc}}{\partial x_\gamma} \frac{1}{2|\overline{E}_{inc}|} \frac{\partial E_\chi^{inc}}{\partial x_\chi} \right)_{surf}$$

(91)

Here $T$ - are sums of transformed contributions, which depend on the transformed dipole and quadrupole moments and the indices $(s, p)$.

## 8. SELECTION RULES FOR THE CONTRIBUTIONS

In accordance with (83) the cross-section is expressed via the sum of the contributions $T_{f_1 - f_2 - f_3}$, which depend on the three dipole and quadrupole moments transforming after irreducible representations of the symmetry group. In accordance with the expressions (75-82) these contributions are not equal to zero when

$$C_{V_{(s,p)}} [f_1, f_2, f_3] \neq 0 \tag{92}$$

Let us consider one, the first line in the first sum in (72). The condition

$$R_{n,l,(s,p)} \langle n|f_1|m\rangle \langle m|f_2|r\rangle \langle r|f_1|l\rangle \neq 0 \tag{93}$$

is valid, when the following conditions

$$R_{n,l,(s,p)} \neq 0 \tag{94}$$

$$\langle r|f_1|l\rangle \neq 0 \tag{95}$$

$$\langle m|f_2|r\rangle \neq 0 \tag{96}$$

$$\langle n|f_3|m\rangle \neq 0 \tag{97}$$

are fulfilled. Let us designate the irreducible representations which determine transformational properties of the $(s, p)$ vibration and of the dipole and quadrupole moments $f$ by the symbol $\Gamma$. The expression (94) is valid when

$$\Gamma_{(s,p)} \in \Gamma_l \Gamma_n, \tag{98}$$

while the other expressions can be used for determination of the irreducible representations for the wavefunctions in their right side,

$$\Gamma_l \in \Gamma_r \Gamma_{f_1} \quad , \quad \Gamma_r \in \Gamma_m \Gamma_{f_2} \quad \text{and} \quad \Gamma_m \in \Gamma_n \Gamma_{f_3} \tag{99}$$

After the consecutive substitution of the expressions (99) in (98) one can obtain the following condition, when (93) is valid

$$\Gamma_{(s,p)} \in \Gamma_{f_1} \Gamma_{f_2} \Gamma_{f_3} \tag{100}$$

Analysis of another lines in (72) results in the same expression (100). Thus the expression (100) presents selection rules for the $T_{f_1-f_2-f_3}$ contributions.

Here we have obtained the selection rules, which are valid in the symmetry groups, pointed out above. The condition was that the dipole and quadrupole moments $Q_{\alpha\beta}$ $(\alpha \neq \beta)$ transform after irreducible representations. For the case of the groups where the dipole and quadrupole moments $Q_{\alpha\beta}$ transform after reducible representations one can transfer to the combinations of these moments, transforming after irreducible representations. Then after transformation of the cross-section (74), similar to those, made with the moments $Q_{\alpha\alpha}$, one can obtain the expression for the SEHRS cross-section that coincides formally with (83) with the $T_{f_1-f_2-f_3}$ contributions which slightly differ from those obtained for the previous case. All the $f_1, f_2$ and $f_3$ moments are now combinations of the dipole and quadrupole moments, which transform after irreducible representations of the corresponding symmetry groups and the selection rules (100) are valid now for all point groups. The definition of the main and minor quadrupole moments remains the same and is determined by the constancy or changeability of the sign under the symmetry operations, while the definition of the main and minor dipole moments depends on many factors and is determined by the orientation of the major part of the adsorbed molecules with respect to the enhanced component of the electric field $E_z$, which is perpendicular to the surface.

## 9. QUALITATIVE CLASIFICATION OF THE CONTRIBUTIONS AFTER ENHANCEMENT DEGREE

In accordance with our previous consideration the most enhancement for the strongly rough surface is caused by the quadrupole interaction with the main quadrupole moments, which we shall designate as $Q_{main}$ and by the dipole interaction with the main dipole moments $d_{main}$. Than the contributions $T_{f_1-f_2-f_3}$ for the monolayer coverage or less, which we shall designate further simply as $(f_1-f_2-f_3)$ can be classified qualitatively after the enhancement degree in the following manner:

1. $(Q_{main}-Q_{main}-Q_{main})$ - the most enhanced scattering type.

2. $(Q_{main}-Q_{main}-d_z)$ - scattering type, which can be strongly enhanced too, but in a lesser degree than the previous one.

3. $(Q_{main}-d_z-d_z)$ - scattering type, which can be strongly enhanced too, but lesser, than the two previous ones and

4. $(d_z-d_z-d_z)$ - scattering type, which can be strongly enhanced too, but lesser than the three previous ones.

Here and further we mean under $(f_1-f_2-f_3)$ all contributions with permutations of the $f$ moments. Considering molecules with $C_{nh}, D$ and higher symmetry one can note, that the first and the third enhancement types contribute to the lines, caused by vibrations transforming after the unit irreducible representation or by the totally symmetric vibrations, while the second and the forth types contribute to the lines, caused by vibrations transforming such as the $d_z$ moment. Thus the most enhanced lines in molecules with $C_{nh}, D$ and higher symmetry are caused by the above types of vibrations. The first ones are forbidden in usual HRS. Thus these lines must be an essential feature of the SEHR spectra of symmetrical molecules with the above symmetry. The above consideration deals with the molecules with some definite orientation of the molecule with respect to the surface, when the chosen orientation of the $d_z$ moment of the molecule coincides with the direction of the $E_z$ component of the electric field, which is perpendicular to the surface. However sometimes molecules can be arbitrary oriented at the surface, because of their superposition. Then all the $d$ moments can contribute to the scattering. Thus the contributions of $(Q_{main} - Q_{main} - d_\alpha)$ and $(Q_{main} - d_\alpha - d_\beta)$ scattering types ($\alpha, \beta = x, y, z$), can manifest in the SEHRS spectra. The other contributions without $Q_{main}$, or those, which contain $Q_{\min or}$ moments apparently will be small and may manifest in the SEHR spectra only in very strong incident field. Now in order to demonstrate correctness of the selection rules and our classification, let us interpret the SEHRS spectra of pyrazine, phenazine, and pyridine on the base of the presented theory.

## 10. ANALYSIS OF THE SEHRS SPECTRUM OF PYRAZINE

Analysis of the experimental SEHR spectrum of pyrazine obtained in [16,17] (Figure 4)

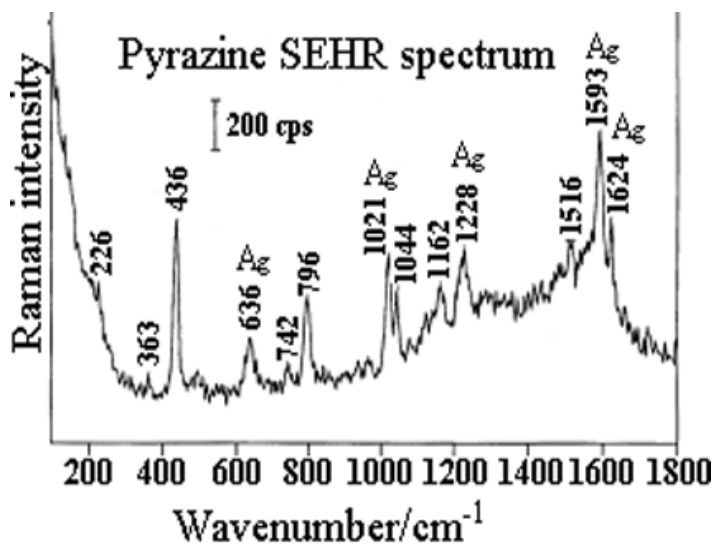

Figure 4. The SEHRS spectrum of pyrazine

is also an interesting confirmation of the dipole-quadrupole theory, since it allows also to explain appearance of the strong forbidden lines with $A_g$ symmetry, which are forbidden in the usual dipole theory. The SEHR spectra, obtained in [16,17] slightly differ one from another because of some difference in experimental conditions. However we have precise information about symmetry of various SEHR lines, which is sufficient for the analysis of the spectra. Below we shall analyze the SEHR spectrum obtained in [16].

Regretfully the authors of [16] did not pointed out some experimental details, which are essential for determination of precise values of wavenumbers and are necessary for the analysis of a real experimental situation. It concerns of the number of adsorbed layers of pyrazine. Apparently there can be several adsorbed layers in this experiment. In this case the pyrazine molecules can adsorb in three manners because of possible superposition of the molecules: flatly, arbitrary and vertically binding via the nitrogen atom with the surface. Due to these possible orientations all the $d$ moments can be essential for the scattering. Before the interpretation of the pyrazine spectrum, it is necessary to define orientation of the molecule with respect to the coordinate system. In accordance with the results [16,17] the $z$ axis passes via the two nitrogen atoms and the $y$ axis is perpendicular to the plane of the molecule. In accordance with the selection rules (100), the lines caused by the vibrations with the following symmetry are observed in the SEHR spectrum (Table 1):

1. $A_g$ - (636, 1021, 1228, 1593 and 1624 $cm^{-1}$) caused mainly by $(Q_{main} - Q_{main} - Q_{main})$ and $(Q_{main} - d_\alpha - d_\alpha)$ scattering contributions of vertically, flatly, and arbitrary adsorbed pyrazine. These lines are forbidden in usual HRS and their appearance strongly proves our point of view.

2. $B_{1u}$ - 1044 $cm^{-1}$ caused mainly by the $(Q_{main} - Q_{main} - d_z)$ and $(d_\alpha - d_\alpha - d_z)$ scattering contributions of vertically and arbitrary adsorbed pyrazine. Apparently the contributions $(Q_{main} - Q_{main} - d_z), (d_y - d_y - d_z)$ of horizontally adsorbed pyrazine are small since they contain the $d_z$ moment, which is associated with the non enhanced $E_y$ tangential component of the electric field for this orientation.

Table 2 Symmetry and the wavenumbers of observable lines of the SEHR spectrum of pyrazine. Here vs – very strong, s – strong, m – middle, w- weak, vw – very weak.

| Symmetry type | Wavenumbers for SEHRS ($cm^{-1}$) | Relative intensity |
|---|---|---|
| $A_g$ | 1624 | m |
| $A_g$ | 1593 | vs |
| $A_g$ | 1228 | s |
| $A_g$ | 1021 | s |
| $A_g$ | 636 | s |
| $B_{1u}$ | 1044 | m |
| $B_{2u}$ | 796 | s |
| $B_{2u}$ | 436 | vs |
| $B_{3u}$ | 1162 | m |
| $B_{2g}$ | 1516 | w |
| $B_{3g}$ | 742 | w |

3. $B_{2u} - 436$ and 796 $cm^{-1}$ caused mainly by the $(Q_{main} - Q_{main} - d_y)$ and $(d_y - d_y - d_y)$ scattering contributions of horizontally and arbitrary adsorbed pyrazine, when the $d_y$ moment is parallel to the enhanced $E_z$ component of the electric field, which is perpendicular to the surface, or has a projection on this direction. The vertically adsorbed pyrazine apparently does not determine the intensity of the lines with $B_{2u}$ symmetry, since the corresponding contributions $(Q_{main} - Q_{main} - d_y)$ and $(d_z - d_z - d_y)$ include the non enhanced tangential component of the electric field $E_y$.

4. $B_{3u}$ - 1162 $cm^{-1}$ caused mainly by the $(Q_{main} - Q_{main} - d_x)$ and $(d_y - d_y - d_x)$ scattering contributions of arbitrary adsorbed pyrazine. Apparently the contributions $(Q_{main} - Q_{main} - d_x)$ and $(d_y - d_y - d_x)$ of vertically and horizontally adsorbed pyrazine are small, since they contain the non enhanced component of the field $E_x$.

5. The lines with the $B_{2g}$ and $B_{3g}$ symmetry (1516 and 742 $cm^{-1}$ respectively) may be caused mainly by $(Q_{main} - d_z - d_x)$ and $(Q_{main} - d_z - d_y)$ scattering contributions of arbitrary oriented molecules. The absence of the lines of $A_u$ and $B_{1g}$ symmetry caused by $(d_z - d_x - d_y)$, $(Q_{main} - d_x - d_y)$ and similar scattering contributions may be caused by small enhancement of these lines in the spectrum due to the non enhanced components of the electric field, which define the intensity of these contributions.

Thus the appearance of all the lines in SEHRS on pyrazine can be explained by our theory, while appearance of the strong lines with $A_g$ symmetry strongly confirms existence of the strong quadrupole light-molecule interaction.

## 11. ANALYSIS OF THE SEHRS SPECTRUM OF PHENAZINE

The geometry of the phenazine molecule with the $D_{2h}$ symmetry group is shown on Figure 5.

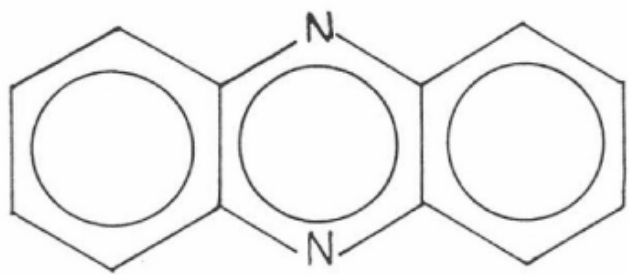

Figure 5. The geometry of the phenazine molecule.

Let us consider the SEHRS spectrum of phenazine, obtained in [15], which is presented on Figure 6 [11,12,14]. The orientation of the molecule relative to the coordinate system we shall chose as for the pyrazine molecule. The $z$ axis passes via two nitrogen atoms and the $y$ axis is perpendicular to the plane of the molecule. The main difficulty in this matter is assignment of vibrations of phenazine to various irreducible representations which are not well defined due to a large size of the molecule and as a consequence a small distance between the frequencies of various vibrations and also very inaccuracy of determination of their values. Here we use the assignment from [25]. It appears that some modes can be assigned to various irreducible representations due to close values of calculated wavenumbers which may belong to various irreducible representations and are used for identification. This fact prevents to unequivocal assignment of these modes. Assignment of the SEHRS lines of phenazine for 0 V and -0.2 V applied potentials is taken from [25] (Table 2).

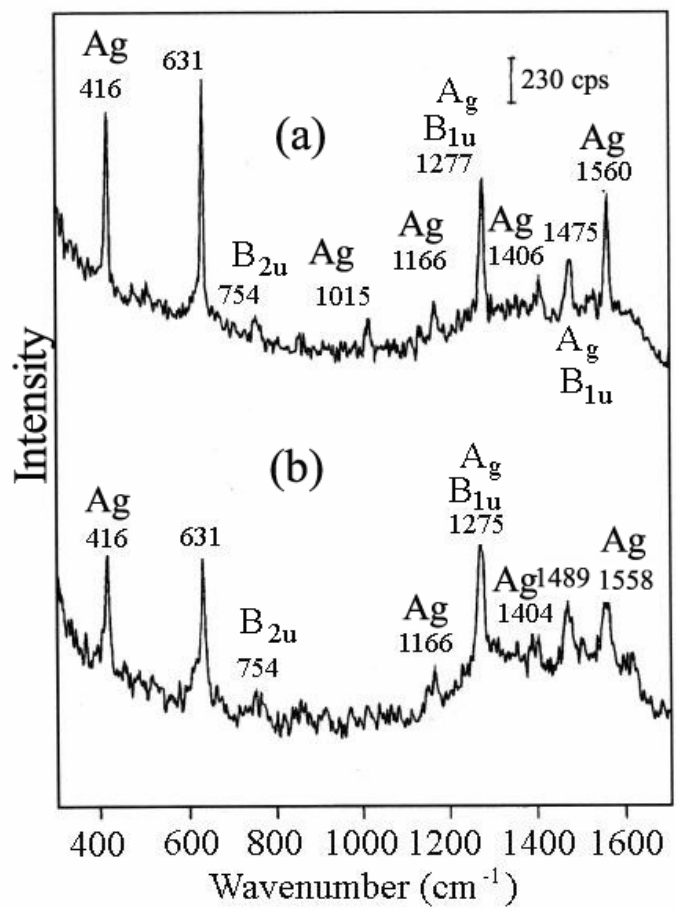

Figure 6. The SEHRS spectra of phenazine for **a**) 0 V  **b**) -0.2 V potentials.

Table 2. Symmetry and the wavenumbers of observable lines of the SEHR spectrum of phenazine. The values of wavenumbers in parenthesis correspond to those in [25].

| SEHRS | | Assignment | | | |
|---|---|---|---|---|---|
| 0 V $(cm^{-1})$ | -0.2 V $(cm^{-1})$ | $A_g$ | $B_{1u}$ | $B_{2u}$ | $B_{3u}$ |
| 416s | 416s | $A_g$ | | | |
| 631vs | 631vs | | $B_{1u}$ (657) | | |
| 754w | 754vw | | | $B_{2u}$ (749) | |
| 1015w | | $A_g$ | | | |
| 1166w | 1166w | $A_g$ | | | |
| 1277s | 1275s | $A_g$ (1280) | $B_{1u}$ (1275) | | |
| 1406w | 1404w | $A_g$ | | | |
| 1475m | 1469m | $A_g$ (1479) | $B_{1u}$ (1471) | | |
| 1560s | 1558m | $A_g$ | | | |

The main feature of the spectrum is appearance of sufficiently strong lines with $A_g$ symmetry which are forbidden in usual HRS. They are the lines with 416, 1015, 1166, 1406 and 1560 $cm^{-1}$ (here we write out the wavenumbers, which refer to the 0 V applied potential). This result is in a rigorous agreement with our theory. The enhancement of these lines is caused by the contributions $(Q_{main} - Q_{main} - Q_{main})$ and $(Q_{main} - d_\alpha - d_\alpha)$, of the molecules with arbitrary orientation. Here we take into account that all the dipole moments are the main ones for an arbitrary orientation.

The other two lines at 1277 $cm^{-1}$ and 1475 $cm^{-1}$ may refer both to the $A_g$ and $B_{1u}$ symmetry types, because of the impossibility of the rigorous assignment of these lines due to the close values of the wavenumbers, which are used for the assignment [25]. However, in both cases they can be explained by the $(Q_{main} - Q_{main} - Q_{main})$ and $(Q_{main} - d_\alpha - d_\alpha)$ scattering types for the $A_g$ symmetry and by the $(Q_{main} - Q_{main} - d_z)$ and $(d_\alpha - d_\alpha - d_z)$ scattering types of the arbitrary oriented molecules for the $B_{1u}$ symmetry type. Both types of the lines should be strongly enhanced and therefore they manifest in the SEHR spectrum of phenazine. The line 754 $см^{-1}$, caused by the vibrations with $B_{2u}$ symmetry type can be explained by the $(Q_{main} - Q_{main} - d_y)$ and $(d_\alpha - d_\alpha - d_y)$ scattering types of the arbitrary oriented molecules. Thus existence of all lines of phenazine, observed in [15] can be explained by the dipole-quadrupole theory.

## 12. ANALYSIS OF THE SEHRS SPECTRUM OF PYRIDINE

The SEHRS spectra of pyridine are presented in [8,9,16,17]. These spectra differ slightly one from another that can be explained by various experimental conditions in these works. However, the positions of the observed SEHRS lines are nearly the same for all these spectra. Therefore we shall analyze the spectrum, obtained in [9] (Figure. 7). The pure dipole approximation is able to explain main peculiarities of the SEHR spectrum of pyridine. This fact is associated with specific symmetry of this molecule and with the fact that the $d_z$ moment of the molecule transforms in accordance with the unit irreducible representation of the $C_{2v}$ symmetry group. Here the z axis passes via the nitrogen atom and the carbon atom which is opposite to the nitrogen atom, while the y axis is perpendicular to the plane of the molecule. It is the reason that all the observed lines are allowed in the dipole approximation. Therefore, pyridine is not the molecule which succeeds in discovery of the strong quadrupole light-molecule

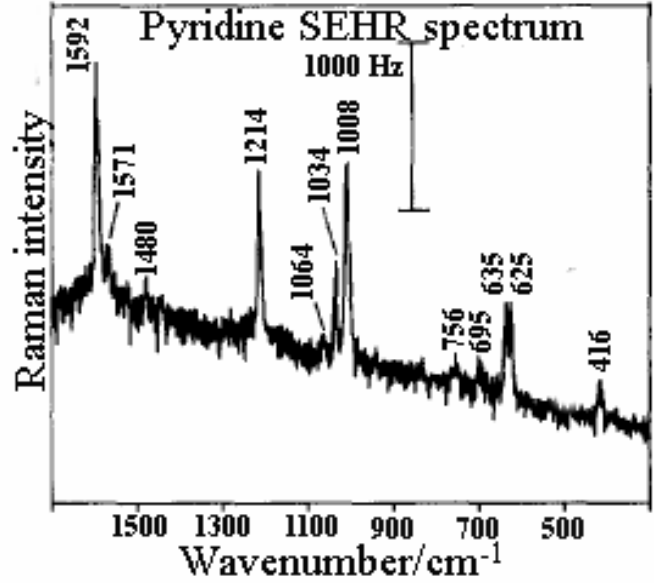

Figure 7. The SEHR spectrum of pyridine [9].

interaction. However, we can reconsider the SEHRS results of on the base of the dipole-quadrupole theory and explain all the features of the SEHR spectra. Firstly, one should note that the calculated wavenumbers of the pyridine vibrations, obtained in [9] differ strongly from the values, observed experimentally from the SEHR spectrum. This result is apparently associated with the use of very approximate methods of determination of the vibrational wavenumbers in [9]. However, the experimental wavenumbers well fit the values, published in [26] for the SERS spectrum of pyridine on $Ag$ with 0.2L exposure. Therefore, we use the values, published in [26] for the symmetry analysis. The pyridine molecule adsorbs primary vertically, binding with the surface via nitrogen. However, a part of the molecules can adsorb horizontally, or in arbitrary manner because of possible overlaying of the molecules even in the first adsorbed layer. Then in accordance with our theory all the $Q_{xx}, Q_{yy}, Q_{zz}, d_x, d_y, d_z$ moments may be essential for the scattering and can be the main moments for this system. However the influence of the $d_x$ moment apparently is significantly less than of the moments $d_y$ and $d_z$, since superposition of two pyridine molecules do not change their plane orientation significantly. Since $Q_{xx}, Q_{yy}, Q_{zz}$ and $d_z$ moments transform after the unit irreducible representation, in accordance with the selection rules (100), the most enhanced contributions

$(Q_{main} - Q_{main} - Q_{main}), (Q_{main} - Q_{main} - d_z), (Q_{main} - d_\alpha - d_\alpha)$ and

$(d_\alpha - d_\alpha - d_z)$, $\alpha = (x, y, z)$, define strongest enhancement of the lines, caused by the totally symmetric vibrations with $A_1$ symmetry (Table 3).

Table 3. Symmetry and wavenumbers of observable lines of the SEHR spectrum of pyridine.

| Symmetry | The mode number | Wavenumbers ($cm^{-1}$) [9] | Relative Intensity |
|---|---|---|---|
| $A_1$ | 3 | 625 | s |
| | 1 | 1008 | vs |
| | 6 | 1034 | s |
| | 8 | 1064 | vw |
| | 5 | 1214 | vs |
| | 9 | 1480 | vw |
| | 4 | 1592 | vs |
| $B_1$ | 14 | 1571 | wv |
| $B_2$ | 27 | 416 | w |
| | 26 | 695 | w |
| | 23 | 756 | w |

Appearance of the lines caused by vibrations with $B_2$ symmetry is caused mainly by the $(Q_{main} - Q_{main} - d_y)$ and $(d_\alpha - d_\alpha - d_y)$ types of the scattering by horizontally and arbitrary oriented molecules. In this case the $d_y$ moment, which is perpendicular to the plane of pyridine possesses a component which is parallel to the enhanced component of the electric field $E_z$ and the scattering contributions pointed become large. Appearance of the lines, caused by the vibrations with $B_1$ symmetry, is apparently caused by $(Q_{main} - Q_{main} - d_x)$ and $(d_\alpha - d_\alpha - d_x)$ from the molecules superposed one another which lie nearly flatly on the surface. Since the primary orientation of the adsorbed pyridine is vertical, the number of other orientations apparently is small and the intensity of the lines with $B_1$ and $B_2$ symmetry is small too.

One should note that there is a sufficiently strong line at $635 cm^{-1}$ in the experimental SEHR spectra of pyridine, published in [8,9]. This line is absent in the pyridine spectra in [16, 17] and among the calculated wavenumbers for pyridine in [8,9]. Therefore, we do not discuss the origin of this line in the paper. Thus, after the analysis of the SEHR spectrum of pyridine we can assert that the dipole-quadrupole theory successfully explains its SEHR spectrum.

## 13. SOME NOTIONS ABOUT INTERPRETATION OF THE SEHRS SPECTRUM OF TRANS-1, 2-BIS (4-PYRIDYLE) ETHYLENE

Another molecule, which was used for investigation of SEHRS [7] is trans-1, 2-bis (4-pyridyle) ethylene with $C_{2h}$ symmetry group. The main feature of the analysis in [7], that it was performed only in the dipole approximation. In addition the authors of [7] use very approximate numerical methods for calculations of various characteristics of Hyper Raman processes like vibrational wavenumbers (frequencies), polarizabilities, hyperpolarizabilities, their normal coordinate derivatives, line intensities and some others. From our point of view the neglecting by the strong quadrupole light molecule interaction and the use of these methods result in a very large discrepancy of the experimental results on these spectra and their calculating values. Let us consider the results, which refer to trans-1, 2-bis (4-pyridyle) ethylene in the adsorption geometry, when the long $y$ axis of the molecule is perpendicular to the surface. In accordance with the above SEHRS theory only the lines with $A_u$ and $B_u$ symmetry can be observed in the dipole approximation.

Table 4. Calculated and experimental wavenumbers of the SER and SEHR lines of trans-1, 2-bis (4-pyridyle) ethylene and their symmetry types in accordance with [7].

| The numbers of vibrations of $A_g$ symmetry | Calculated wavenumbers of $A_g$ symmetry $(cm^{-1})$ | The numbers of vibrations of $B_u$ symmetry | Calculated wavenumbers of $B_u$ symmetry $(cm^{-1})$ | Experimental wavenumbers for SEHRS of $B_u$ symmetry $(cm^{-1})$ | Experimental wavenumbers for SERS of $A_g$ symmetry $(cm^{-1})$ |
|---|---|---|---|---|---|
| 17 | 280 | 16 | 466 | 552 | 320 |
| 16 | 645 | 15 | 537 | 599 | 652 |
| 15 | 675 | 14 | 677 | 688 | 663 |
| 14 | 866 | 13 | 814 | 841 | 847 |
| 13 | 986 | 12 | 987 | 972 | 1008 |
| 12 | 1071 | 11 | 1071 | 1007 | 1064 |
| 11 | 1097 | 10 | 1095 | 1116 | |
| 10 | 1145 | 9 | 1137 | 1198 | 1200 |
| 9 | 1201 | 8 | 1221 | 1208 | 1200 |
| 8 | 1224 | 7 | 1239 | 1289 | 1244 |
| 7 | 1340 | 6 | 1302 | 1325 | 1314 |
| 6 | 1363 | 5 | 1366 | 1341 | 1338 |
| 5 | 1409 | 4 | 1418 | 1422 | 1421 |
| 4 | 1498 | 3 | 1504 | 1489 | 1493 |
| 3 | 1560 | 2 | 1569 | 1548 | 1544 |
| 2 | 1609 | 1 | 1610 | 1593 | 1604 |
| 1 | 1682 | | | | 1640 |
| | 3018 | | 3012 | | |
| | 3034 | | 3033 | | |
| | 3049 | | 3049 | | |
| | 3061 | | 3061 | | |
| | 3069 | | 3069 | | |
| The numbers of vibrations of $B_g$ symmetry | Calculated wavenumbers of $B_g$ symmetry $(cm^{-1})$ | The numbers of vibrations of $A_u$ symmetry | Calculated wavenumbers of $A_u$ symmetry $(cm^{-1})$ | Experimental wavenumbers for SEHRS of $A_u$ symmetry $(cm^{-1})$ | Experimental wavenumbers for SERS of $B_g$ symmetry $(cm^{-1})$ |
| 8 | 409 | 9 | 298 | 306 | 400 |
| 7 | 503 | 8 | 408 | 385 | 491 |
| 6 | 747 | 7 | 572 | 664 | 738 |
| 5 | 829 | 6 | 761 | 803 | 802 |
| 4 | 900 | 5 | 865 | 878 | 881 |
| 3 | 956 | 4 | 901 | | 955,972 |
| 2 | 1029 | 3 | 1004 | | 1064 |
| 1 | 1047 | 2 | 1033 | 1060 | |
| | | 1 | 1047 | | |

This result corresponds to the results and conditions published in [7]. However consideration of the calculated vibrational wavenumbers of the lines with the $A_g$ and $B_u$ symmetry demonstrates nearly the same values for major vibration wavenumbers within the calculation errors (0-10) $cm^{-1}$ (Table 4). ). For example this refer to the vibrations with $A_g$ and $B_u$ symmetry under the numbers ((15-14), (13-12), (12-11), (11-10), (10-9), (8-8), (6-5), (5-4), (4-3), (3-2), (2-1), and by all others in fact. Here the first number refers to the vibrations with $A_g$ and the second with $B_u$ symmetry. In this situation assignment of the experimental SEHR lines to $A_g$ and $B_u$ symmetry types is impossible, since the uncertainty in determination of the calculated wavenumbers of the lines with this symmetry is less then the uncertainty of assignment of the measured and calculated values, which can differ one from another till (50-60) $cm^{-1}$. Thus in accordance with the results of Table 4 many experimental SEHR lines, which were assigned to the $B_u$ irreducible representation can be assigned to the $A_g$ one. One should remind, that in accordance with the regularities of the SEHRS spectra, which were explained theoretically by the dipole-quadrupole theory and were observed experimentally [20], the most enhanced lines are those caused by the totally symmetric vibrations, transforming after the unit irreducible representation $A_g$.

Comparison of a large number of wavenumbers with the $B_u$ symmetry, measured in the SEHRS experiments with the lines of $A_g$ symmetry demonstrates that they differ one from another on a very small value ~ (1-10) $cm^{-1}$, which is less that the calculation and measurement values. This fact refer to the following lines with $A_g$ and $B_u$ symmetry respectively ((847-841), (1008-1007), (1200-1198), (1200-1208), (1338-1341), (1421-1422), (1493-1489), (1544-1548)). Here the wavenumbers are pointed out in $cm^{-1}$. Thus the both above results incline us to opinion, that a large number of the SEHRS lines, which were assigned to the $B_u$ symmetry, can have the $A_g$ symmetry in fact. Taking into account that the lines with $A_g$ symmetry are allowed in the SEHRS spectra of 1, 2-bis (4-pyridyle) ethylene in the dipole-quadrupole theory, the above facts do not contradict to explanation of its spectrum in terms of the strong dipole and quadrupole light-molecule interactions.

Before analyzing of the SEHRS spectrum of 1, 2-bis (4-pyridyle) ethylene it is necessary to consider the real possible orientations of this molecule with respect to the surface. In principle the molecule can adsorb in two main manners-vertically with the main $y$ axis perpendicular to the surface and in flat orientation. Since interaction of the molecule with the surface is weak, one can consider, that the molecule symmetry preserves for both orientations [20]. The enhancement of the lines with $A_g$ symmetry is caused mainly by $(Q_{main} - Q_{main} - Q_{main})$, $(Q_{main} - d_z - d_z)$ contributions in the horizontal orientation and by $(Q_{main} - Q_{main} - Q_{main})$ and $(Q_{main} - d_y - d_y)$ in the vertical orientation in accordance with classification of the moments after enhancement degree and the selection rules (100). In fact the intensities of these lines are determined by the contributions of both orientations. The enhancement of the lines with $A_u$ symmetry is caused mainly by the $(Q_{main} - Q_{main} - d_z)$ and $(d_z - d_z - d_z)$ contributions in the horizontal orientation and by $(Q_{main} - Q_{main} - d_z)$ and $(d_y - d_y - d_z)$ contributions in the vertical orientation, where $d_y$ is perpendicular to the surface andis parallel to the $E_z$ component of the electric field, which is perpendicular to the surface. The last contributions apparently are small, because they are caused by $d_z$ moment and by $E_y$ tangential component of the electric field which is not enhanced. The lines with $B_u$ symmetry are caused mainly by $(Q_{main} - Q_{main} - d_{min\,or})$ and $(d_z - d_z - d_{min\,or})$ scattering types, where under $d_{min\,or}$ we mean $d_x$ and $d_y$ components of the dipole moment in the horizontal orientation and mainly by $(Q_{main} - Q_{main} - d_y)$ and $(d_y - d_y - d_y)$ contributions, which can be strongly enhanced in the vertical orientation. In principle the lines with $B_g$ symmetry can appear in the SEHRS spectra too mainly due to existence of the contributions $(Q_{main} - d_z - d_x)$ and $(Q_{main} - d_z - d_y)$ in the horizontal orientation and mainly due to ($Q_{main} - d_y - d_z$) contributions in the vertical orientation. Analysis of experimental SERS and SEHRS lines demonstrates existence of very close values of the wavenumbers of the lines with $B_g$ symmetry in SERS and $A_u$ symmetry in SEHRS (Table 4). This refer to the lines ((802-803), (881-878), (1064-1060)) respectively. The wavenumbers as earlier are pointed out in $cm^{-1}$. Taking into account the large uncertainty in determination of symmetry of the SERS and SEHRS lines due to the calculation errors, one can make a conclusion, that a part of the SEHRS lines can belong to $A_g$ and $B_g$ symmetry. Regretfully we are not able to explain the SEHRS results of [7] more precisely because of the large negligence in calculations of the wavenumbers and in determination of the lines symmetry in this work. However consideration of the strong quadrupole light-molecule interaction and of the two possible orientations of the molecules permits to explain actually the possibility of appearance of the lines with $A_g$ and $B_g$ symmetry in the SEHR spectra of trans-1, 2-bis (4-pyridyle) ethylene.

### 14. SOME NOTIONS ABOUT THE SEHRS SPECTRA OF CRYSTAL VIOLET

One of symmetrical molecule, which was used for investigation of the SERS and SEHRS spectra is the crystal violet (CV) molecule. The most reliable determination of the symmetry of vibrations of this molecule was made in [27] where one considered that its symmetry group is $D_3$. However the authors point out the irreducible representation $A$, in spite of there is no such irreducible representation in this symmetry group. There are only the irreducible representations $A_1$ and $A_2$. Apparently the authors meant the irreducible representation $A_1$. In [28] the spectra of infrared absorption and the SERS and SEHRS spectra were investigated partly. However the authors consider that the molecule symmetry group is $C_3$. Recently the authors of [29,30] investigated the spectra of SERS, SEHRS, of the usual RS, infrared absorption and the usual HRS, considering that the group of the molecule is $D_3$, such as in [27]. In addition they made their own assignment on the base of the DFT and B3LYP-311G methods. Careful consideration of experimental and calculated vibrational lines, vibrational wavenumbers, and possible assignment published in all these papers, reveal large discrepancy in calculated values and assignment of the lines (see [27-30]). Therefore our attempts to make our own assignment of the SEHRS lines failed in fact and we consider, that the matter of precise assignment is spurious now. However there are several lines, which assigned to the $A_1$ irreducible representation in all these works. They are the lines at 1621 and 1389 $cm^{-1}$. In addition comparison of the IR, SER and SEHR spectra, published in [28] points out that the lines 1298, 1370, 1478 and 1588 $cm^{-1}$ are observed in all these spectra. Therefore in accordance with selection rules in IR spectra these lines can be caused only by vibrations with $A_2$ and $E$ symmetry. Consideration of the SER and SEHR spectra in accordance with the dipole-quadrupole theory [14, 20] demonstrates, that these lines may be caused preferably by the $(Q_{main} - d_z)$ and $(Q_{main} - Q_{main} - d_z)$ contributions in SERS and SEHRS respectively. This refer first of all to the lines at 1370 and 1589 $cm^{-1}$ because of their large relative intensity in the SERS and SEHRS spectra. The assignment of these lines to the $E$ symmetry may also be possible since the lines may be caused by the $(Q_{main} - (d_x, d_y))$ and $(Q_{main} - Q_{main} - (d_x, d_y))$ contributions, for molecules, superposed in the first and second layers having orientation which is not parallel to the surface. As it was considered earlier this fact depends on the coverage of substrate which is not known in [28]. Thus apparently there is a multilayer coverage in these experiments. Appearance and existence of the lines with $A_1$ symmetry in the SEHR spectra [28-30] is in agreement with our theory. One should note, that in the $D_3$ symmetry group, appearance of the lines, caused by vibrations with all possible irreducible representations in SEHRS formally is possible in a pure dipole theory, such as in pyridine. However appearance of the lines with $A_2$ symmetry in SERS can be explained only by the $(Q_{main} - d_z)$ type of the scattering.

Therefore the correct explanation of the SEHR spectra of CV should be made on the base of the dipole-quadrupole theory, which is able to explain all other features of the SER and SEHR spectra of CV also. One should note, that the SEHR spectra of CV in the above mentioned works sometimes were obtained with the incident radiation, when the resonance scattering is possible. The resonance character of these processes must not change the selection rules for the contributions [100]. Therefore the result, that the lines, caused by vibrations with $A_1$ and $A_2$ symmetry are observed should be valid in the case of surface enhanced resonance Raman scattering and strongly support our theory.

## 15. ELECTRODYNAMICAL FORBIDDANCE AND PECULIARITIES OF THE SEHRS SPECTRA OF THE METHANE MOLECULE

In the methane molecule, which belongs to the $T_d$ symmetry group the dipole and quadrupole moments with various indices transform after irreducible representation. However the quadrupole moments with the same indices transform after reducible representations. The combinations of the quadrupole moments, transforming after irreducible representation $Q_1, Q_2$ and $Q_3$ have the form

$$Q_1 = \frac{1}{3}(Q_{xx} + Q_{yy} + Q_{zz}) \tag{101}$$

$$Q_2 = \frac{1}{2}(Q_{xx} - Q_{yy}) \tag{102}$$

$$Q_3 = \frac{1}{4}(Q_{xx} + Q_{yy} - 2Q_{zz}) \tag{103}$$

and

$$Q_{xx} = Q_1 + \frac{2}{3}Q_3 + Q_2 \tag{104}$$

$$Q_{yy} = Q_1 + \frac{2}{3}Q_3 - Q_2 \tag{105}$$

$$Q_{zz} = Q_1 + \frac{2}{3}Q_3 - 2Q_2 \tag{106}$$

Then one can obtain the following expression for $|\overline{E}|(\overline{ef}_e)$, which is contained in the light-molecule interaction Hamiltonians (14, 15)

$$|\overline{E}|(\overline{ef}_e) = (\overline{Ed}) + \frac{1}{2}div\overline{E} \times \left(Q_1 + \frac{2}{3}Q_3\right) + \frac{1}{2}\left(\frac{\partial E_x}{\partial x} - \frac{\partial E_y}{\partial y} - 2\frac{\partial E_z}{\partial z}\right)Q_2$$
$$+ \frac{1}{2}\sum_{\substack{\alpha\beta \\ \alpha \neq \beta}} \frac{\partial E_\alpha}{\partial x_\beta} Q_{e\alpha\beta} \tag{107}$$

The value $Q_1$ is the main moment with a constant sign, which transform after the unit irreducible representation and is responsible for the most enhancement. In our case one can see from (107) that there is a factor $div\overline{E} = 0$, which make equal to zero the contributions from this moment. This result is an Electrodynamical forbiddance of the strong quadrupole light-molecule interaction, which realizes in the methane molecule [31]. Due to this forbiddance, associated with the cubic symmetry and with electrodynamical law, all terms, associated with $Q_1$ and $Q_3$ are equal to zero and do not influence on formation of the enhanced SEHRS spectra. Let us consider peculiarities of the SEHRS spectrum of methane. Its SEHRS spectrum must be determined by the most enhanced contributions, where the classification in accordance with the enhancement degree has been made above. However in the methane molecule all the contributions, which contain the main quadrupole moment $Q_{main} = Q_1$ are forbidden due to the Electrodynamical forbiddance. Therefore the first three contributions in this molecule are equal to zero and the most enhanced contribution is $(d_z - d_z - d_z)$. The theoretical group analysis demonstrates, that this contribution transforms after reducible representation, which contain two irreducible representations $F_1$ and $F_2$. Thus the most enhanced lines do not belong to the vibrations with the unit irreducible representation, as it was in molecules with lower symmetry, $D_{2h}$ in particular. The remaining contributions, which may be enhanced due to the presence of the $d_z$ moment, $(d_z - d_z - Q_{\min or})$, $(d_z - d_z - d_{\min or})$, $(d_z - d_{\min or} - d_{\min or})$ can contribute in principle in the lines, caused by the totally symmetric vibrations. However these contributions are significantly weaker than the contribution $(d_z - d_z - d_z)$. Therefore these lines must be very weak, or must be absent at all in the SEHRS spectrum of methane.

One should note, that for the multilayer coverage, the moments $d_x$ and $d_y$ also can refer to the main moments. However as it has been pointed above, because the enhancement in the second and upper layers is significantly weaker than the enhancement in the first layer, the contributions $(d_z - d_x - d_x)$, $(d_z - d_y - d_y)$ and $(d_z - d_x - d_y)$ will be enhanced significantly lower, than the contribution $(d_z - d_z - d_z)$. Therefore even the multilayer coverage must not result in appearance of the strong lines, caused by the totally symmetric vibrations. Taking into account that methane has no vibrations with the $F_1$ symmetry, one obtains the result that the lines with the $F_2$ irreducible representation must dominate in the spectrum.

The Electrodynamical forbiddance of the strong quadrupole light-molecule interaction must be observed in molecules, belonging to another cubic symmetry groups $O, O_h, T,$ and $T_h$, where linear combinations of the quadrupole moments $Q_1, Q_2$ and $Q_3$ have the same form. This must result in the fact, that in another cubic molecules, such as in methane, the lines, caused by the totally symmetric vibrations must not experience the most enhancement. The most enhanced lines will be caused by another irreducible representations. The matter of presence of the lines with the totally symmetric vibrations in the spectra of these molecules is very difficult and is associated with a specific symmetry group of these molecules. From our point of view, this matter must be solved together with experimental investigations of the SEHRS spectra of specific molecules, which are absent at present.

## 16. SOME NOTIONS CONCERNING THE CHEMICAL ENHANCEMENT MECHANISM IN SEHRS AND THE NATURE OF THE FIRST LAYER EFFECT

As it was mentioned above the enhancement in the first adsorbed layer in SEHRS is significantly stronger than the enhancement in the second layer. One connects this phenomenon with the direct contact of the molecules with the substrate, which is the reason of the so-called "chemical enhancement". This effect is discovered and is known in SERS. From our point of view these ideas are absolutely wrong.

As it was demonstrated above, the most enhancement arises in the regions of the maxim curvature of the surface, or on the so-called "active sites", or "hot spots". Let us consider again the roughness of the wedge, cone or spike type. The electric field in the region of the top of such surface is well described by the formulae (11). The relations of the electric fields and their derivatives in the first and the second adsorbed layers can be estimated as

$$\left(\frac{r_2}{r_1}\right)^{6\beta} = (3)^{6\beta} \qquad (108)$$

or

$$\left(\frac{r_2}{r_1}\right)^{6+6\beta} = (3)^{6+6\beta} \qquad (109)$$

respectively. Here $r_1$ and $r_2$ - are the distances from the top of the roughness to the centre of gravity of the molecules (Figure 8). One can see for example that for $\beta \sim 1$ these values change in the interval $\sim 10^3 - 10^5$, that explain completely the difference in the enhancement in the first and the second layers of the adsorbed molecules. Thus the first layer effect is associated with a very strong change of the electric field and its derivatives, when one moves from the surface and has a pure electrodynamical nature. One should note that the fact that the "chemical mechanism" is a mistake idea was first published in the works on SERS in [18-20].

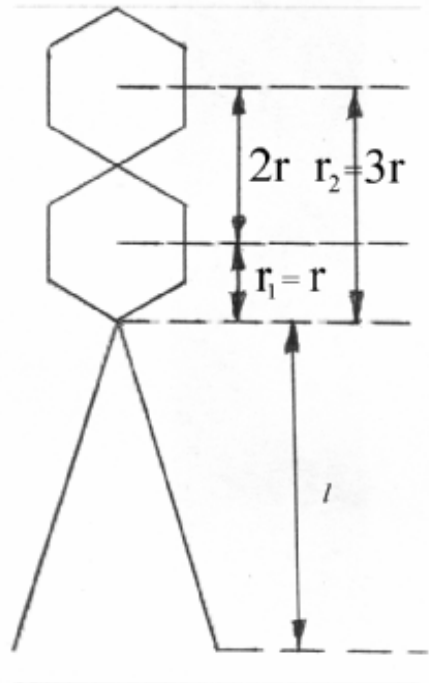

Figure 8. The geometry of adsorption of two layers of molecules near the model roughness

### 17. SOME PECULIARITIES OF THE SEHRS FREQUENCY DEPENDENCE

In [32] it was note that the signal of SEHRS strongly varies in the region of wavelengths of the incident light $\lambda = 750 nm$ and something above. It was pointed out that for $\lambda = 750 nm$ and less the signal is practically absent and strongly increases when the wavelength increases. We would like to point out that such behaviour corresponds to our ideas in full. It is necessary to point out, that for $\lambda = 750 nm$ the dielectric constant of silver is negative and its modulus is sufficiently large that allows to consider that the substrate is ideally conductive. For the wavelength of the scattering light in SEHRS $\lambda \sim 375 nm$ the dielectric constant becomes very small, because these frequencies are close to the plasma frequency of silver. Therefore the metal changes strongly its properties and becomes close to the transparent material. In this case our ideas about ideally conductive medium are not suitable. The enhancement of the scattering field strongly decreases that results in the strong decrease of the signal of SEHRS.